\documentclass[journal]{IEEEtran}

\usepackage{authblk}

\usepackage{framed,multirow}

\usepackage{amssymb}
\usepackage{latexsym}

\usepackage{url}
\usepackage{xcolor}

\usepackage{amsmath}
\usepackage{graphicx}
\usepackage{caption}
\usepackage{subcaption}
\usepackage{wrapfig}
\usepackage{float}
\usepackage{amsthm}
\usepackage{diagbox}
\usepackage{stmaryrd}
\usepackage{comment}
\usepackage{color,soul}
\usepackage{multirow}

\usepackage{makecell}

\usepackage{xcolor,cancel}

\setstcolor{red}

\floatstyle{plaintop}
\restylefloat{table}

\usepackage{hyperref}
\hypersetup{
	colorlinks=true,
	linkcolor=blue,
	filecolor=magenta,      
	urlcolor=cyan,
}

\usepackage[linesnumbered,lined,boxed,commentsnumbered]{algorithm2e}

\begin{document}

\title{Geodesic Density Regression for Correcting 4DCT Pulmonary Respiratory Motion Artifacts}

\author[1,2]{Wei Shao}
\author[1]{Yue Pan}
\author[3]{Oguz C. Durumeric}
\author[4]{Joseph M. Reinhardt}
\author[5]{John E. Bayouth}
\author[2]{Mirabela Rusu}
\author[1,6]{Gary E. Christensen}

\affil[1]{Department of Electrical and Computer Engineering, University of Iowa, Iowa City, IA 52242 USA}
\affil[2]{Department of Radiology, Stanford University, Stanford, CA 94305 USA}
\affil[3]{Department of Mathematics, University of Iowa, Iowa City, IA 52242 USA}
\affil[4]{Roy J. Carver Department of Biomedical Engineering, University of Iowa, Iowa City, IA 52242 USA}
\affil[5]{Department of Human Oncology, University of Wisconsin - Madison, Madison, WI 53792 USA}
\affil[6]{Department of Radiation Oncology, University of Iowa, Iowa City, IA 52242 USA}

\maketitle

\begin{abstract}
Pulmonary respiratory motion artifacts are common in four-dimensional computed tomography (4DCT) of lungs and are caused by missing, duplicated, and misaligned image data.
This paper presents a geodesic density regression (GDR) algorithm to correct motion artifacts in 4DCT by correcting artifacts in one breathing phase with artifact-free data from corresponding regions of other breathing phases.
The GDR algorithm estimates an artifact-free lung template image and a smooth, dense, 4D (space plus time) vector field that deforms the template image to each breathing phase to produce an artifact-free 4DCT scan.
Correspondences are estimated by accounting for the local tissue density change associated with air entering and leaving the lungs, and using binary artifact masks to exclude regions with artifacts from image regression.
The artifact-free lung template image is generated by mapping the artifact-free regions of each phase volume to a common reference coordinate system using the estimated correspondences and then averaging.
This procedure generates a fixed view of the lung with an improved signal-to-noise ratio.
The GDR algorithm was evaluated and compared to a state-of-the-art geodesic intensity regression (GIR) algorithm using simulated CT time-series and 4DCT scans with clinically observed motion artifacts.
The simulation shows that the GDR algorithm has achieved significantly more accurate Jacobian images and sharper template images, and is less sensitive to data dropout than the GIR algorithm.
We also demonstrate that the GDR algorithm is more effective than the GIR algorithm for removing clinically observed motion artifacts in treatment planning 4DCT scans.
Our code is freely available at \url{https://github.com/Wei-Shao-Reg/GDR}.
\end{abstract}

{\bf Keywords: }{Geodesic regression, artifact correction, motion artifact, 4DCT, image registration, lung cancer}


\section{Introduction}
\begin{figure}[!hbt]
	\centering
	\includegraphics[height=1.2in]{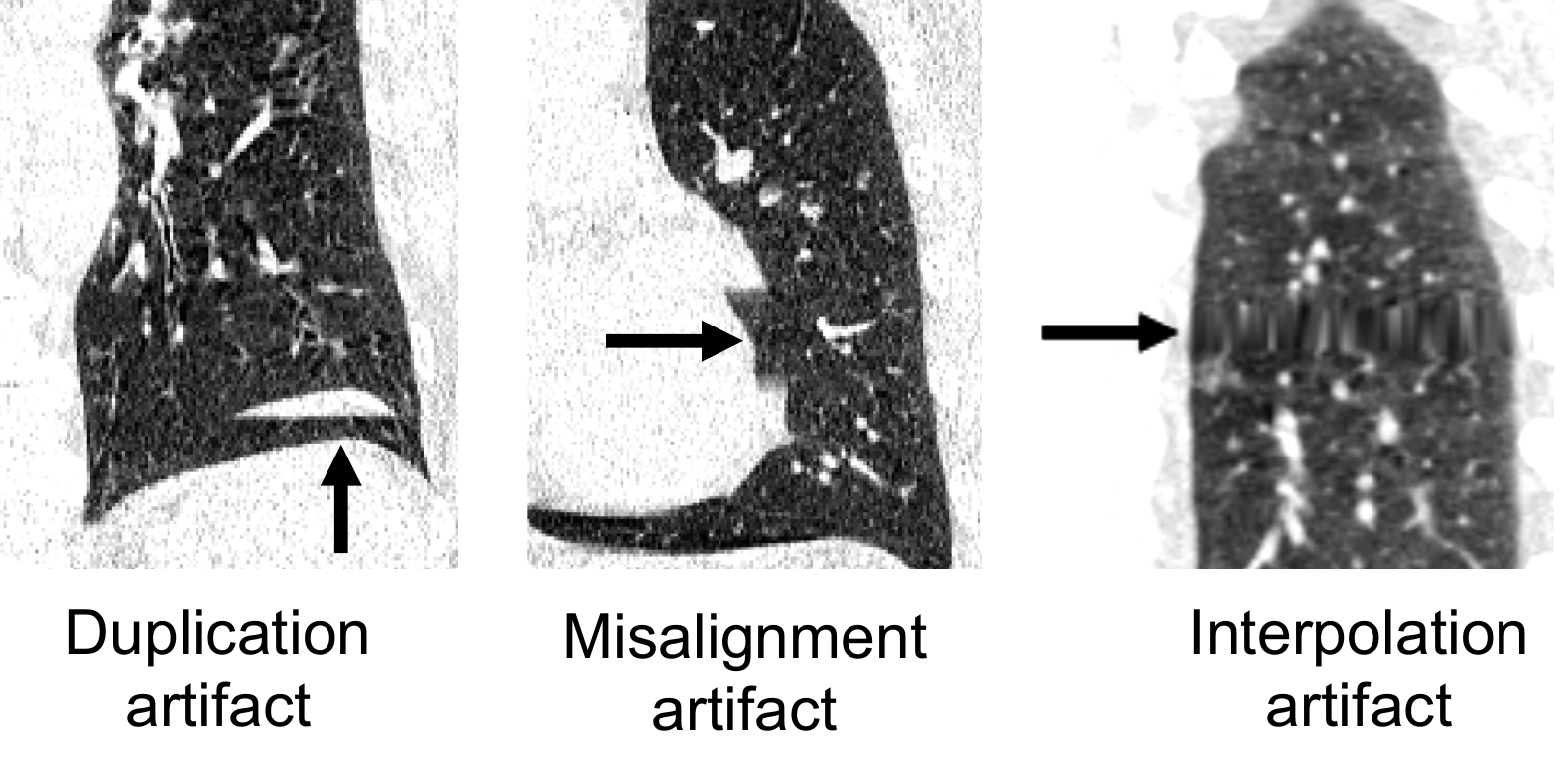}
	\caption{Arrows show common 4DCT motion artifacts. Interpolation is used to estimate missing data.
		\label{fig:intro_common_artifacts}}
\end{figure}

Four-dimensional computed tomography (4DCT) scans are routinely acquired of lung cancer patients for radiation treatment planning.
The quality of 4DCT often suffers from artifacts, i.e., distortions in lung anatomy representation.
Yamamoto et al.~\cite{YamamotoTokihiro2008RAoA} showed that 90\% of 4DCT scans had at least one artifact near the diaphragm or the heart, with a mean thickness of artifact stacks of 11.6 millimeters (mm).
The most common artifacts that occur when acquiring pulmonary 4DCT images during free breathing are due to irregular breathing.
Common breathing motion artifacts include duplicated structure near the diaphragm (duplication artifact), misalignment of adjacent stacks (misalignment artifact), and interpolation artifact (see Fig.~\ref{fig:intro_common_artifacts}).

\begin{figure}[!hbt]
	\centering
	\includegraphics[height=1.4in]{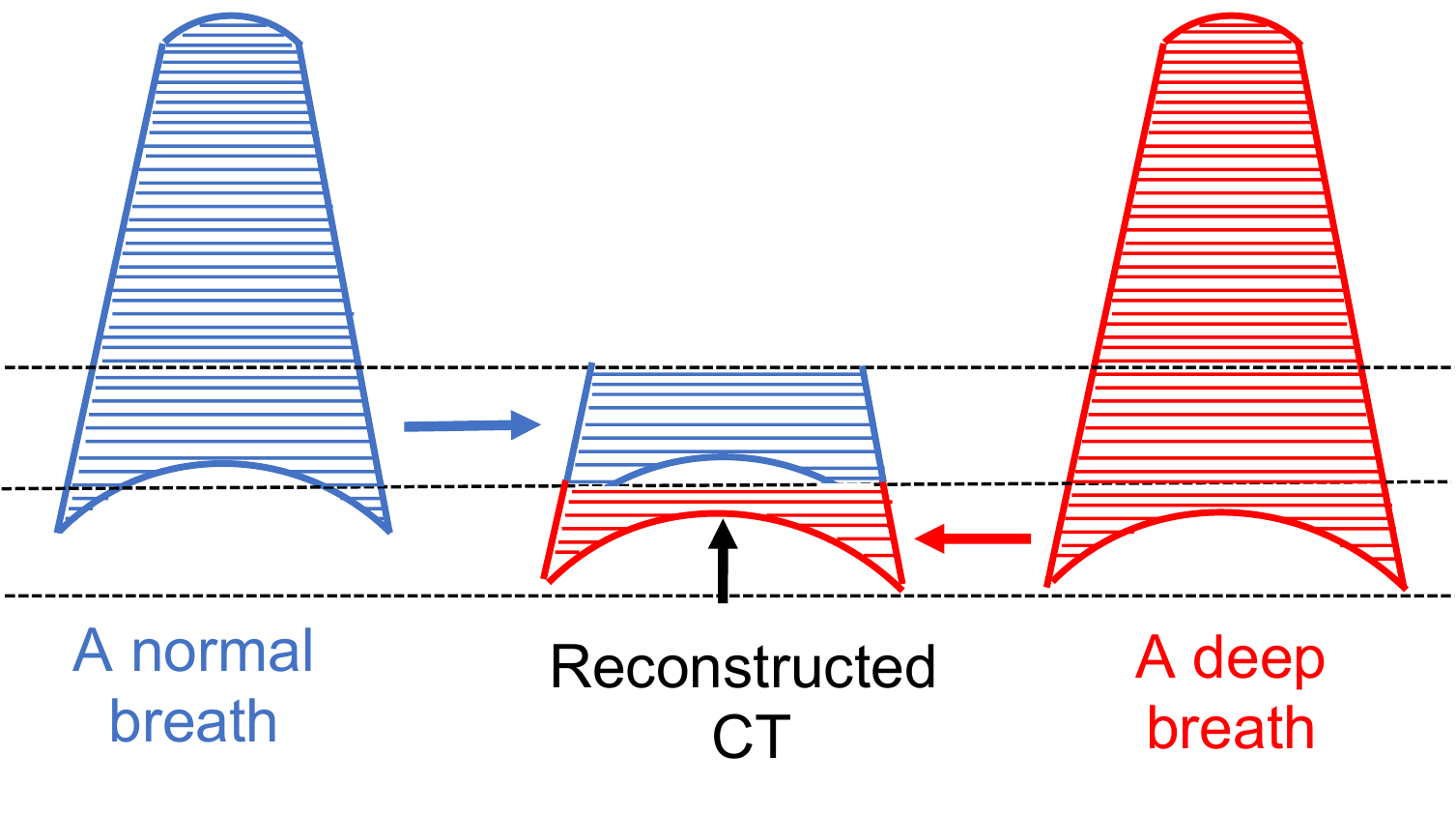}
	\caption{ Illustration of a duplication artifact near the diaphragm caused by a normal breath followed by a deep breath.  
		4DCT image volumes are constructed from multiple breaths since the whole lung cannot be imaged at once.
		A stack of images from one breath (left) is placed on top of a stack of images from the following breath (right) to produce the reconstructed CT image volume (middle).
		\label{fig:intro_duplicate_artifact}}
\end{figure}

4DCT image volumes are constructed by concatenating stacks of transverse slices from scans collected over multiple breathing cycles since the field of view of a CT scanner does not cover the entire height of the lung. 
Irregular breathing causes motion artifacts at stack boundaries when consecutive scans image the same region of the lung or miss a region of the lung partially or entirely.
Figure~\ref{fig:intro_duplicate_artifact} shows a phantom example illustrating how irregular breathing can cause a duplication artifact.

Motion artifacts in 4DCT can affect the quality of  4DCT-guided radiation therapy (RT), which is commonly used to treat lung cancer patients.
Functional avoidance RT improves conventional RT by using Jacobian images derived from image registration of 4DCT to minimize doses to high-function lung tissues~\cite{Patton2018}.
However, motion artifacts in 4DCT often cause errors in image registration and the corresponding Jacobian images~\cite{Shao2018Sensitivity}, degrading the accuracy of functional avoidance RT.
Numerous approaches have been developed to reduce motion artifacts in cine and helical 4DCT scans.
Approaches designed for reduction of motion artifacts in cine 4DCT include generating interpolated CT scans exactly at the predefined tidal volumes based on an optimal flow method~\cite{Ehrhardt2007}, optimizing phase or displacement-based sorting protocols by incorporating anatomic similarity~\cite{JohnstonEric2011R4Ca}, and estimating respiratory motion based on principle component analysis of the motion fields
generated from an artifact-free reference CT image~\cite{ZhangYongbin2013Mrmf}.
Current geodesic regression algorithms may mitigate motion artifacts in both cine and helical 4DCT scans by averaging artifact data over all respiratory phases~\cite{miller2006geodesic,fletcherRegression2011,NiethammerRegression2011,HinkleJacob20124Cir,SinghVectorMomenta2013,FISHBAUGH20171,hong2017fast}.
Those regression algorithms are either designed for shape regression or image regression that assumes constant voxel intensity in response to deformation.

This paper presents a geodesic density registration (GDR) algorithm to remove motion artifacts that occur in a 4DCT image during breathing.
The GDR algorithm estimates a smooth 4D transformation that estimates the respiratory motion of the lung during breathing.
This transformation is used to pullback all phases of the 4DCT image to the coordinate system of the reference phase to create an intensity averaged template image.
The GDR algorithm accommodates image intensity change associated with breathing by using a tissue density deformation action and is therefore more suitable for regression of pulmonary images than algorithms that assume constant image intensity~\cite{miller2006geodesic,fletcherRegression2011,NiethammerRegression2011,HinkleJacob20124Cir}.
In the most common case, motion artifacts occur at the same location in the reference coordinate system in some of the pullback phase images, but not in all phases.
The goal of the GDR algorithm is to exploit this fact to only use good data to estimate the 4D transformation and to compute the average intensity template image.
Binary masks drawn in each phase of the 4DCT image volume are used to identify and exclude artifact regions when estimating the 4D motion transformation and when computing the average intensity image.
We present for the first time the mathematical equation to pullback the masks when using a tissue density preserving costs function.
This paper also shows how Pontryagin’s Minimum Principle (see Appendix~\ref{sec:pmp}) can be used to solve the image regression problem.

This paper provides experimental evidence that not every artifact needs to be masked for the GDR approach to reduce the effects of motion artifacts.
The reason for this is that the regression process averages the effects of good and bad data which helps mitigate errors caused by motion artifacts.
Drop out experiments are presented which show that as much as 50\% of the 4DCT image can be masked out and still have a reasonably good GDR image registration result.
These two observations taken together imply that artifact masks do not have to be accurately identified to provide benefit.
Thus, artifact masks can be quickly identified by hand (See Section~\ref{sec:mask}) or automatically identified by a computer using a deep learning approach. 

\section{Methods}
\subsection{Data Acquisition}
\label{sec:4DCTdata}

This institutional review board-approved study used 30 lung cancer patients from the University of Wisconsin Health University Hospital.
For each subject, two 4DCT scans were acquired before radiation therapy using a Siemens EDGE CT scanner in helical mode with 120 kVp, 100 mAs per rotation, tube rotation period slightly higher than 0.5 seconds, 0.09 pitch, 76.8 mm beam collimation, 128 detector rows, and a reconstructed slice thicknesses of 0.6 mm. All CT volumes were then resampled isotopically to $1\times 1 \times 1 \text{mm}^3$ in this study.
We used voice instruction during the scanning to guide the patients to have more regular breathing~\cite{Bayouth_muscial_2019}.
Each 4DCT scan consists of 10 breathing phase 3D image volumes, labeled 20\% (20IN), 40\% (40IN), 60\% (60IN), 80\% (80IN) and 100\% (100IN) inspiration and 80\% (80EX), 60\% (60EX), 40\% (40EX), 20\% (20EX) and 0\% (0EX) expiration.

\subsection{Geodesic Image Regression}
\begin{figure}[!hbt]
	\centering
	\begin{subfigure}[t]{0.22\textwidth}
		\centering
		\includegraphics[height=0.75in]{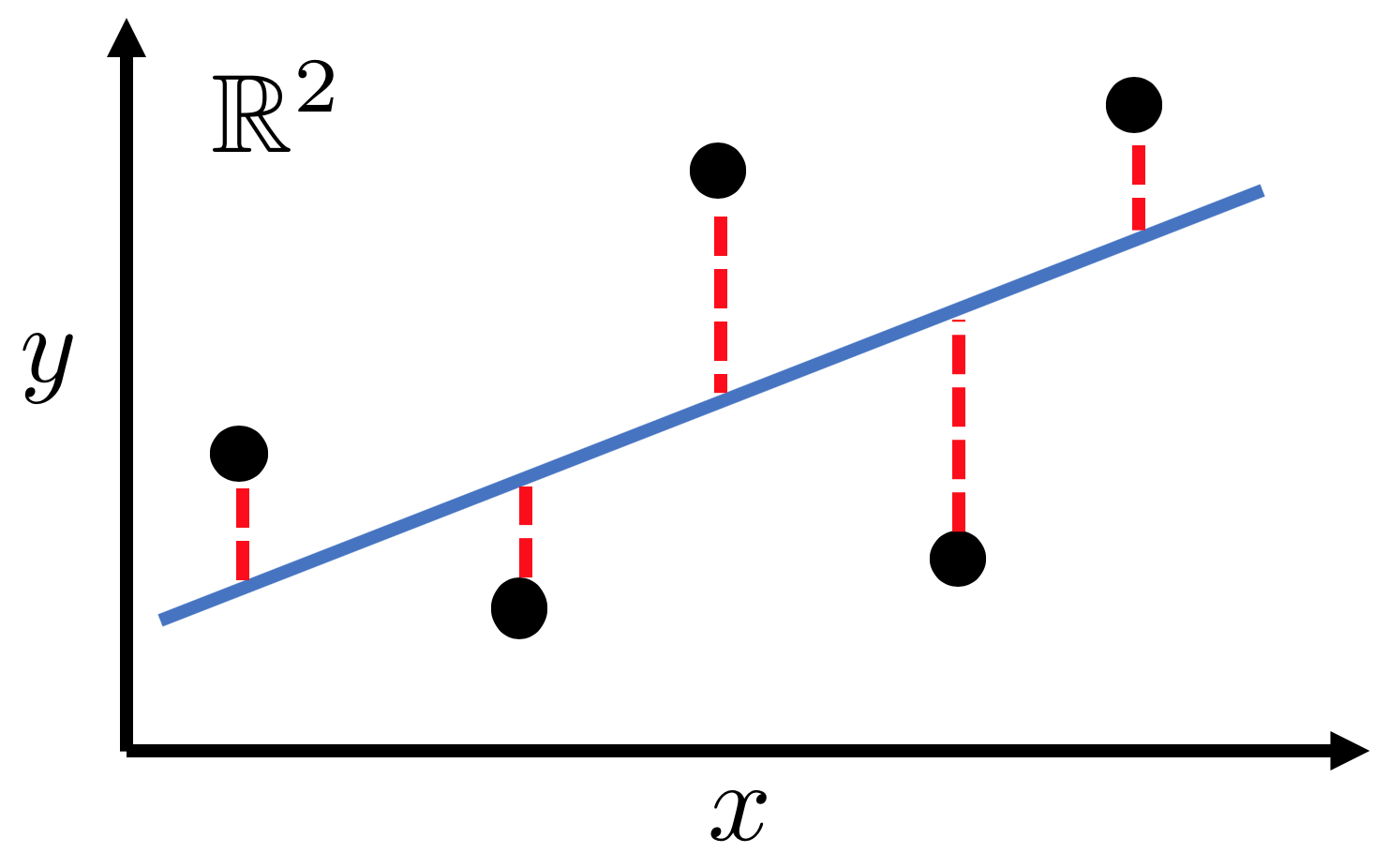}
		\caption{Simple linear regression}
		\label{fig:linearReg}
	\end{subfigure}
	\hspace{0.1in}
	\begin{subfigure}[t]{0.2\textwidth}
		\centering
		\includegraphics[height=0.75in]{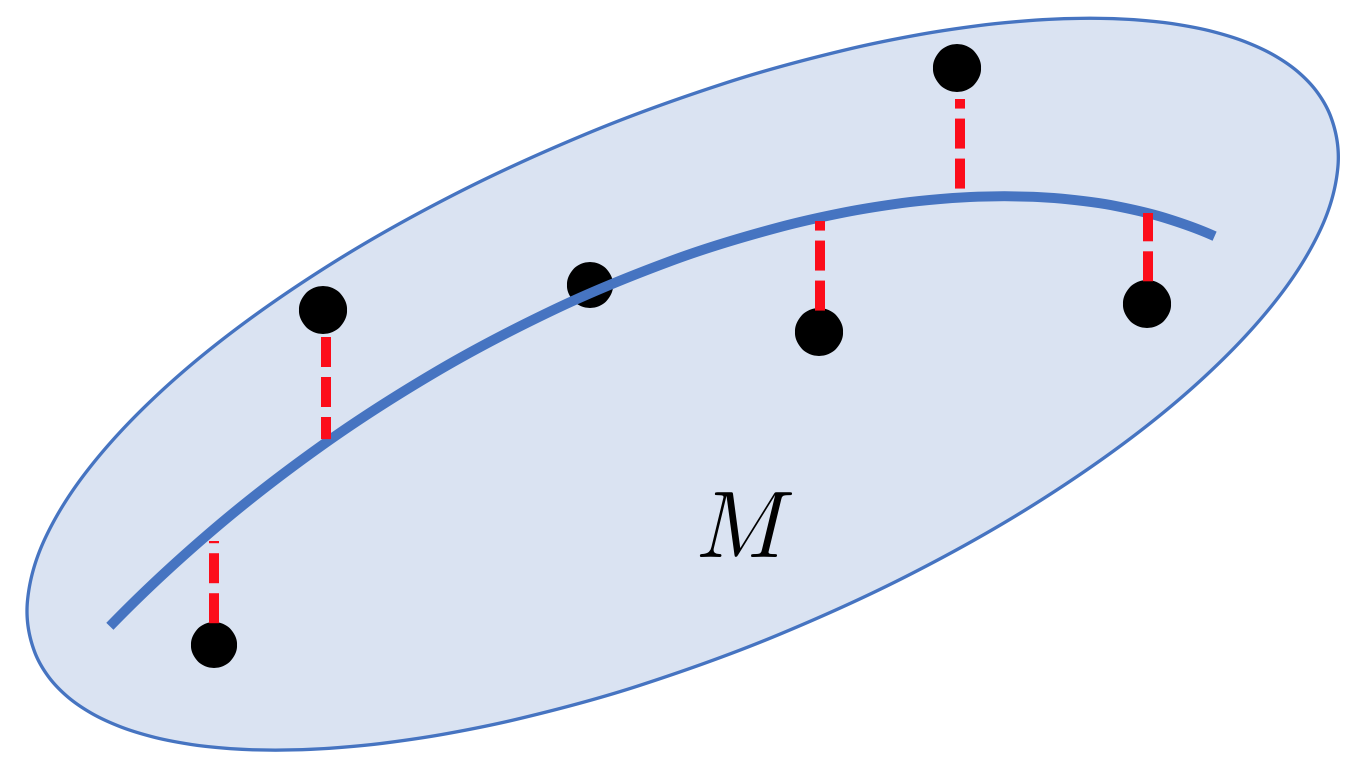}
		\caption{Geodesic image regression}
		\label{fig:geoReg}
	\end{subfigure}
	\caption{Simple linear regression and geodesic image regression. In panel (a), a black point represents a point in the 2D Euclidean space. In panel (b), a black point represents an image in the Riemannian manifold $M$ of all smooth images with compact support.
		\label{fig:linearReg_vs_geoReg}}
\end{figure}

The simplest form of regression analysis is the simple linear regression (see Fig.~\ref{fig:linearReg}).
Given $n$ observations $\{(x_i,y_i),i=0,\cdots,N-1\}$ in the 2D plane $\mathbb{R}^2$, the goal is to find an optimal straight line $y = \alpha + \beta x$ that describes the linear relationship between the independent variable $x$ and the dependent variable $y$.
The simple linear regression problem can be solved by minimizing the sum of squared differences cost function $C$
\begin{equation}
C(\alpha,\beta) = \sum_{i=0}^{N-1} d_E^2(y_i,\alpha + \beta x_i)
\end{equation}
where $d_E$ is the one-dimensional Euclidean distance.

Geodesic image regression (see Fig.~\ref{fig:geoReg}) is a generalization of the simple linear regression to a Riemannian manifold~\cite{fletcherRegression2011,NiethammerRegression2011}.
In geodesic image regression, time $t$ is the independent variable and image $I$ is the dependent variable. 
Suppose at each time $t_i$, there is an observation $I_i$ from the Riemannian manifold $M$ of admissible images.
The goal of geodesic image regression is to find a template image $I_T$ and a flow of diffeomorphisms $\phi_t$ that minimizes the following cost function
\begin{equation}
C(I_T,\phi_t) = \sum_{i=0}^{N-1} d_M^2\big(I_i, I(t_i)\big)
\end{equation}
where $N$ is the number of input images, $\{I_i\}$ are the input images, $I(t) \triangleq \phi_t \cdot I_T$ is a flow of deformed template images, and $d_M$ is the Riemannian distance on $M$.

\subsection{Review of the Large Deformation Model}
Traditionally, an image $J:\mathbb{R}^3\rightarrow\mathbb{R}$ is deformed by transformations in the diffeomorphism group $\mathcal{G} = Diff(\mathbb{R}^3) \triangleq  \{\varphi :\mathbb{R}^3 \rightarrow \mathbb{R}^3 \hspace{1mm}|\hspace{1mm}\varphi \hspace{1mm}\text{is a diffeomorphism}\}$  via the left group action $\varphi \cdot J \triangleq J\circ(\varphi^{-1}) = J(\varphi^{-1})$~\cite{beg2005-LDDMM}.
Given two images, the goal of diffeomorphic image registration is to find a diffeomorphism $\varphi$ that deforms one image into the shape of the other image. 
In the large deformation diffeomorphic metric mapping (LDDMM) setting~\cite{beg2005-LDDMM}, $\varphi = \phi_1$ is the endpoint of a flow of diffeomorphisms $\phi_t:[0,1]\rightarrow \mathcal{G}$ parameterized by a time-varying velocity field $v_t:[0,1]\rightarrow \mathfrak{g}$ via the following ordinary differential equation (O.D.E)~\cite{beg2005-LDDMM,christensen-rabbitt-miller:fluidpaper}
\begin{equation}
\frac{d}{dt}\phi_t(x) = v_t(\phi_t(x))
\label{eq:flow_of_diffeo}
\end{equation}
where $\mathfrak{g}$ is a reproducing kernel Hilbert space of smooth, compactly supported vector fields. 

To guarantee the existence of solutions to the above O.D.E in the space of diffeomorphisms~\cite{dupuis1998,trouve1995}, the Hilbert space $\mathfrak{g}$ is equipped with an inner product $<,>_{\mathfrak{g}}$ given  by
\begin{equation}
<Ka,b>_\mathfrak{g} = <a,b>_{L^{2}} 
\label{eq:sobolev_norm}
\end{equation}
where the operator $K$ is often a Gaussian smoothing kernel.

The diffeomorphism group $\mathcal{G}$ can be considered as a Lie group~\cite{miller2006geodesic,holm2009geometric,bruveris2015geometry}.
The Lie algebra $\mathfrak{g}$ of $\mathcal{G}$ is the tangent space of $\mathcal{G}$ at the identity map, i.e., $\mathfrak{g} = T_{Id}\mathcal{G}$.
Each element $v\in\mathfrak{g}$ is a smooth vector field on $\mathbb{R}^3$ and we assume an inner product $<\cdot,\cdot>_{\mathfrak{g}}$ on $\mathfrak{g}$ is already given by Eq.~\ref{eq:sobolev_norm}.
This inner product $<\cdot,\cdot>_{\mathfrak{g}}$ induces a right-invariant Riemannian metric $g$ on $\mathcal{G}$.
For any $\psi \in \mathcal{G}$, $u_\psi,u'_\psi\in T_\psi \mathcal{G}$, the Riemannian metric $g$ is defined as
\begin{equation}
g_\psi(u_\psi,u'_\psi) = <(D_{\psi}R_{\psi^{-1}})u_\psi,(D_{\psi}R_{\psi^{-1}})u'_\psi>_\mathfrak{g}
\end{equation}
where $R_{\psi^{-1}}$ is the right translation by $\psi^{-1}$, defined as $R_{\psi^{-1}}(\phi) = \phi \circ \psi^{-1}$ for any $\phi \in \mathcal{G} $, $D_{\psi}R_{\psi^{-1}}$ is the push-forward (differential) of $R_{\psi^{-1}}$ at $\psi$ to push the tangent vectors $u_\psi,u'_\psi$ to the Lie algebra $\mathfrak{g}$.

Suppose a flow of diffeomorphisms $\phi_t\in\mathcal{G}$ is parameterized  by a time-varying velocity field $v_t\in\mathfrak{g}$ using Eq.~\ref{eq:flow_of_diffeo}.
Since $(D_{\phi_t}R_{\phi_t^{-1}}) \frac{d\phi_t}{dt} =
\frac{d}{ds}\Big(\phi_s\circ \phi_t^{-1}\Big) \Big|_{s=t} = v_t$,
the kinetic energy of $\phi_t$ measured by the above right-invariant Riemannian metric $g$ is given by
\begin{equation}
E(\phi_t) = \frac{1}{2}\int_{0}^{1}<\frac{d\phi_t}{dt},\frac{d\phi_t}{dt}>_{g_{\phi_t}} = \frac{1}{2}\int_{0}^{1}<v_t,v_t>_\mathfrak{g} dt.
\label{eq:smothness}
\end{equation}

\subsection{Tissue Density Deformation Action}
\begin{figure}[!hbt]
	\centering
	\includegraphics[height=1.6in]{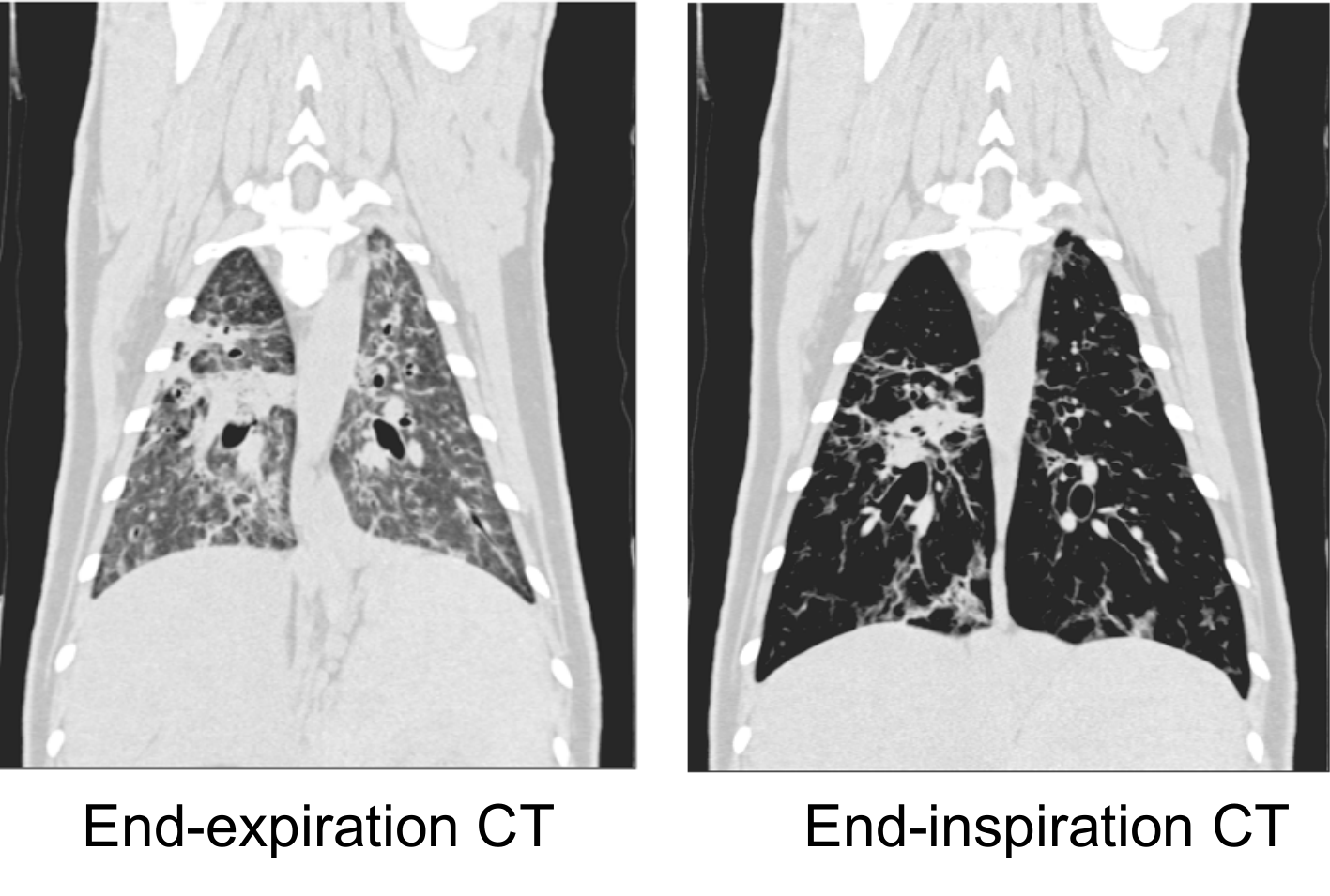}
	\caption{Significant CT intensity change of a sheep lung from the end-expiration phase to the end-inspiration phase. Intensity changes in the human lungs are similar.
		\label{fig:intensity_change}}
\end{figure}

CT intensity in Hounsfield units (HU) varies during respiration because of local tissue density change with the response to deformation of the lungs (see Fig.~\ref{fig:intensity_change} for an example).
Therefore, using the sum of squared CT intensity differences as the similarity cost is not optimal for 4DCT image regression~\cite{yin2009,gorbunova2012mass,cao2010spie}.
To account for variations in CT intensity, we align tissue density images instead of CT intensity images in our regression algorithm.
A CT image in HU can be converted into a tissue density image by
\begin{equation}
I =    \frac{\text{CT} - \text{HU}_{air}}{\text{HU}_{tissue} - \text{HU}_{air}} = \frac{\text{CT} + 1000}{1055}.
\end{equation}
The action of a transformation $\phi$ on a tissue density image $I$ is defined as:~\cite{yin2009,gorbunova2012mass,Bauer2015}
\begin{equation}
\phi \cdot I := |D\phi^{-1}|I\circ(\phi^{-1})
\label{eq:density_action}
\end{equation}
where $|D\phi^{-1}|$ is the Jacobian determinant of $\phi^{-1}$ and is used to account for local tissue density change associated with local lung volume change.

\subsection{Proposed Geodesic Tissue Density Regression Framework}
Given a time-series of lung tissue density images  $I_0, I_1, \cdots, I_{N-1}$ acquired at times $0=t_0<t_1<\cdots<t_{N-2}<t_{N-1}=1$, the goal of geodesic density regression (GDR) is to find a template image $I_T$ and a time-varying velocity field $v_t\in\mathfrak{g}$ to minimize the following cost function
\begin{equation}
E(v_t,I_T) = E(\phi_t) + \frac{1}{\gamma^2}\sum_{i=0}^{N-1}||(\phi_{t_i}\cdot I_T - I_i)M_i||_{L_2}^2
\label{eq:reg_cost}
\end{equation}
where $E(\phi_t)$ is defined in Eq.~\ref{eq:smothness}, $\phi_t$ is parameterized by $v_t$ using Eq.~\ref{eq:flow_of_diffeo},  $\phi_{t_i}\cdot I_T$ is the tissue density deformation action defined in Eq.~\ref{eq:density_action}, and $M_i$ is a binary artifact mask used to exclude artifact regions in the image $I_i$. 
The intensity of the artifact mask $M_i$ is zero in artifact regions and one elsewhere. 
Introducing the artifact masks $M_i$ has several advantages: artifact regions do not contribute to the cost function, data with artifacts regions are not used for template estimation, and thus the proposed algorithm is more robust to CT artifacts.

The tissue density image regression problem can be considered as an optimal control problem.
The time-varying velocity field $v_t$ is considered as the control variable and the flow of images $I(t)$ given by
\begin{equation}
I(t) = \phi_t\cdot I_T = |D\phi_t^{-1}|I_T\circ \phi_t^{-1}
\label{eq:density_flow}
\end{equation}
is considered as the state variable.

The relationship between the control variable and the state variable is given by the following O.D.E.: (see Appendix~\ref{sec:derivative_of_density_flow})
\begin{equation}
\dot{I}(t) \triangleq  \frac{d}{dt}I(t) = -\text{div}\big(I(t)v_t\big)
\label{eq:state_equation}
\end{equation}
with the initial condition $I(0) = I_T$.

Our optimal control problem is to find an admissible control $v_t$ that causes the state $I(t)$ to follow an admissible trajectory that minimizes the performance measure in Eq.~\ref{eq:reg_cost}. 
We introduce the costate variable $\lambda(t)$ and define the following Hamiltonian
\begin{equation}
\begin{split}
H
&= \frac{1}{2}<v_t,v_t>_{\mathfrak{g}} - <\lambda(t),\text{div}\big(I(t)v_t\big)>_{L^2}\\
&= \frac{1}{2}<v_t,v_t>_{\mathfrak{g}} + <I(t),\nabla \lambda(t)\cdot v_t>_{L^2}\\
&= \frac{1}{2}<v_t,v_t>_{\mathfrak{g}} + <I(t)\nabla\lambda(t),v_t>_{L^2}
\end{split}
\label{eq:hamiltonian_4dct}
\end{equation}
where $\lambda(t)$ has the same dimension as $I(t)$.

We use the Pontryagin's Minimum Principle (see Appendix~\ref{sec:pmp}) to solve our regression problem.
The necessary conditions for an optimal solution are given by
\begin{equation}
\begin{cases}
\dot{I}(t) = -\text{div}\big(I(t)v_t\big)\\
\dot{\lambda}(t) = -\nabla \lambda(t)\cdot v_t\\
v_t = -K*\big(I(t)\nabla \lambda(t)\big)
\end{cases}
\label{eq:flow_equations}
\end{equation}
for all $t\in[t_0,t_{N-1}]$, where $K$ is a Gaussian and $*$ is the convolution operation.
The boundary conditions are given by
\begin{equation}
\begin{cases}
I(0) = I_T \\
\lambda(t_{N-1}) = \frac{2}{\gamma^2}\big(I(t_{N-1}) - I_{N-1}\big)M_{N-1}\\
\lambda(t_i^{-}) = \lambda(t_i^{+}) + \frac{2}{\gamma^2}\big(I(t_i) - I_i\big)M_i
\end{cases}
\label{eq:boundary_conditions}
\end{equation}
for all $i\in\{1,\cdots,N-2\}$.

As shown in Appendix~\ref{sec:update_template}, we derived a closed-form formula for computing the template image at each iteration
\begin{equation}
I_T(y) = \frac{\sum_{i=0}^{N-1} (I_i M_i)\circ\phi_{t_i}(y)}{\sum_{i=0}^{N-1} (|D\phi_{t_i}^{-1}|M_i)\circ\phi_{t_i}(y) }.
\label{eq:template}
\end{equation}

Compared to the formula for computing the template image in the previous geodesic intensity regression (GIR) algorithm~\cite{NiethammerRegression2011,SinghVectorMomenta2013}
\[
I_T^{\text{GIR}}(y) = \frac{\sum_{i=0}^{N-1} |D\phi_{t_i}|(y)I_i\circ\phi_{t_i}(y)}{\sum_{i=0}^{N-1} |D\phi_{t_i}|(y) },
\]
our GDR algorithm only uses artifact-free CT data for template estimation and it also accommodates CT intensity change due to respiration.
Notice that the denominator of Eq.~\ref{eq:template} is a scalar image which would be zero in regions where all the pulled back artifact masks are zero.
	Therefore, the GDR image regression may fail if information in one region was excluded from regression in all breathing phases, causing a division-by-zero problem.
	We handle the division-by-zero problem by defining $\frac{0}{0} \triangleq 0$ for Eq.~\ref{eq:template}, i.e., the intensity of a voxel is set to be zero if it is identified as an artifact in all phases.

A simple interpretation of Eq.~\ref{eq:template} is that it computes the average image of all the masked phase images of the 4DCT data set in the coordinate system of $I_T$.
This average is computed using intensity corrected images and accounts for how many non-artifact intensity values are averaged at each voxel.
More specifically, the numerator of Eq.~\ref{eq:template} computes the average of all the artifact-free regions of the 4DCT phase images in the coordinate system of image $I_T$.
This is accomplished by multiplying each phase image by a mask image to exclude regions of the images that contain artifacts. 
Then, each masked image $(I_i M_i)$ is pulled back (mapped) to the common coordinate system of image $I_T$ by the transformation $\phi_{t_i}$ and summed at each voxel location $y$.
The denominator of Eq.~\ref{eq:template} computes the normalizing factor for the average at the voxel location $y$.
This is accomplished by pulling back (mapping) the artifact mask in the coordinate system of each breathing phase to the coordinate system of $I_T$ and averaging the values.
The weighting factor $|D\phi_{t_i}^{-1}|$ takes into account the local CT intensity change due to breathing. 
Thus, an important observation is that Eq.~\ref{eq:template} accounts for both the number of intensity values used in the average and for the intensity changes related to breathing.

\subsection{FoI and FoT GDR Algorithms}
The GDR algorithm may be implemented as a flow-of-images (FoI) or as a flow-of-transformations (FoT).
The FoI GDR algorithm is implemented as an iterative gradient descent algorithm.
The functional derivative of the cost function Eq.~\ref{eq:reg_cost} with respect to the velocity field $v_t$ used in the gradient descent is given by
\[
\nabla_vE(t) = v_t + K*\big(I(t)\nabla \lambda(t)\big)
\]
where $K = \frac{1}{\sigma \sqrt{2\pi}}e^{-\frac{x^2}{2\sigma^2}}$ is a Gaussian smoothing kernel and $\sigma$ is the kernel size.

The iterative FoI GDR algorithm is stated in Algorithm~\ref{al:FoI_GDR}. 
The inputs to the algorithm are the observed image volumes $I_0,\cdots,I_{N-1}$ and the artifact mask image volumes $M_0,\cdots,M_{N-1}$.
The other inputs are the weight $\gamma$ of the data matching term, the Gaussian function $K$ that controls the spatial smoothness of the velocity field, the stopping criteria threshold $\epsilon$, the step size $s$, and the number of past updates $m$.
The FoI GDR algorithm produces an estimate of the static lung $I_T$ at time $t=0$ and the flow of transformations $\phi_t$ that deforms $I_T$ from $t=0$ to $t=1$.
The instantaneous velocity $v_t$, the costate $\lambda(t)$ and time dependent image $I(t)$ are discretized into $P$ computational-time points for computation of steps 2-7.
Suppose we have $k$ computational-time points between any two adjacent image observations, then $P = kN$.

The FoI GDR algorithm can be converted to the FoT GDR algorithm in the following manner. 
Replace $\dot{I}(t) = -\text{div}\big(I(t)v_t\big)$ in step 2 by $I(t) = |D\phi_t^{-1}|I_T\circ \phi_t^{-1}$, and replace $\dot{\lambda}(t) = -\nabla \lambda(t)\cdot v_t$ by $\lambda(t) = \lambda(t_i^-)\circ\phi_{t_i}\circ\phi_{t}^{-1}$, for $t\in [t_{i-1}^+,t_i^-]$ in step 3.
The inverse flow of diffeomorphisms $\phi_t^{-1}$ can be solved using $\frac{d}{dt}{\phi_t^{-1}} = -D\phi_t^{-1} v_t$.  The resulting FoT GDR algorithm is stated in Algorithm~\ref{al:FoT_GDR}.
Note that the inputs and outputs of the FoT and the FoI regression algorithms are the same. 
The FoT algorithm requires three times more computation and storage than the FoI algorithm.

\begin{algorithm*}
	\SetKwInOut{Input}{Input}\SetKwInOut{Output}{Output}\SetKwInOut{Initialization}{Initialization}
	\Input{$I_0,\cdots,I_{N-1}$, $M_0,\cdots,M_{N-1}$, $\gamma$, $\sigma$, $\epsilon$}
	\Output{$I_T$, $v_t$, $\phi_t$}
	\Initialization{$v =0$, $I_T = I_0$, $K = \frac{1}{\sigma \sqrt{2\pi}}e^{-\frac{x^2}{2\sigma^2}}$, $iter = 0$, $m = 3$}
	\BlankLine
	\While{$\text{cost reduction rate} > \epsilon$}{
		Solve state forward: $\dot{I}(t) = -\text{div}\big(I(t)v_t\big)$, $I(0) = I_T$\;
		Solve costate backward: $\dot{\lambda}(t) = -\nabla \lambda(t)\cdot v_t$,
		$\lambda(t_{N-1}) = \frac{2}{\gamma^2}\big(I(t_{N-1}) - I_{N-1}\big) M_{N-1}$,
		$\lambda(t_i^{-}) = \lambda(t_i^{+}) + \frac{2}{\gamma^2}\big(I(t_i) - I_i\big)M_i$ for $i\in\{0,1,\cdots,N-2\}$\;
	    \uIf{$iter \leq m$}{
       (Gradient descent): $v_t \leftarrow v_t - s\big( v_t + K*\big(I(t)\nabla\lambda(t) \big)\big)$\;
       }
       \Else{
        
           (L-BFGS): Calculate Z using Algorithm~\ref{al:L-BFGS}\;
	  	   Calculate s using the strong Wolfe condition\;
	       Update velocity field: $v_t \leftarrow v_t + s Z$}\;
    	Solve forward map flow: $\frac{d}{dt}\phi_t = v_t(\phi_t)$, $\phi(0) = Id$\;
    	Solve inverse map flow: $\frac{d}{dt}{\phi_t^{-1}} = -D\phi_t^{-1} v_t$,$\phi^{-1}(0) = Id$\;
    	Update template: $ I_T = \frac{\sum_{i=0}^{N-1} (I_i M_i)\circ\phi_{t_i}}{\sum_{i=0}^{N-1} (|D\phi_{t_i}^{-1}|M_i)\circ\phi_{t_i} }$\;
    	iter = iter + 1;
	}
	\caption{Flow-of-images (FoI) GDR. Notice that steps 2, 3, 12 and 13 all have an implicit for-loop that loops $P$ times. All differential equations were solved using the Euler method.}
	\label{al:FoI_GDR}
\end{algorithm*}  

\begin{algorithm*}
	\SetKwInOut{Input}{Input}\SetKwInOut{Output}{Output}\SetKwInOut{Initialization}{Initialization}
	\Input{$I_0,\cdots,I_{N-1}$, $M_0,\cdots,M_{N-1}$, $\gamma$, $\sigma$, $\epsilon$}
	\Output{$I_T$, $v_t$, $\phi_t$}
	\Initialization{$v =0$, $I_T = I_0$, $K = \frac{1}{\sigma \sqrt{2\pi}}e^{-\frac{x^2}{2\sigma^2}}$, $iter = 0$, $m = 3$}
	\BlankLine
	
	\While{$\text{cost reduction rate} > \epsilon$}{
		Solve forward maps: $\frac{d}{dt}\phi_t = v_t(\phi_t)$, $\phi(0) = Id$\;
		Solve inverse maps: $\frac{d}{dt}{\phi_t^{-1}} = -D\phi_t^{-1} v_t$,$\phi^{-1}(0) = Id$\;
		Compute state flow: $I(t) = |D\phi_t^{-1}|I_T\circ \phi_t^{-1}$\;
		Compute costate flow: $\lambda(t_{N-1}) = \frac{2}{\gamma^2}\big(I(t_{N-1}) - I_{N-1}\big)M_{N-1}$\;
		$i = N-1$\;
		\While{$i>1$}{
			$\lambda(t) = \lambda(t_i^-)\circ\phi_{t_i}\circ\phi_{t}^{-1}$, for $t\in [t_{i-1}^+,t_i^-]$\;
			$i = i- 1$\;
			$\lambda(t_i^{-}) = \lambda(t_i^{+}) + \frac{2}{\gamma^2}\big(I(t_i) - I_i\big)M_i$\;
		}
   \uIf{$iter \leq m$}{
    (Gradient descent): $v_t \leftarrow v_t - s\big( v_t + K*\big(I(t)\nabla\lambda(t) \big)\big)$\;
  }
  \Else{
           (L-BFGS): Calculate Z using Algorithm~\ref{al:L-BFGS}\;
	  	   Calculate s using the strong Wolfe condition\;
	       Update velocity field: $v_t \leftarrow v_t + s Z$}\;
		Update template: $I_T = \frac{\sum_{i=0}^{N-1} (I_i M_i)\circ\phi_{t_i}}{\sum_{i=0}^{N-1} (|D\phi_{t_i}^{-1}|M_i)\circ\phi_{t_i} }$
	}

	\caption{Flow-of-transformations (FoT) GDR.}
	\label{al:FoT_GDR}
\end{algorithm*}  

\subsection{FoI Versus FoT GDR}
In this study, we use the FoT GDR algorithm instead of the FoI GDR algorithm for lung image regression.
Theoretically, FoI and FoT regression give the same result. However, in practice, these approaches differ in implementation and performance.
The main drawback of the FoI approach for our application is that the O.D.Es in steps 2 and 3 of Algorithm~\ref{al:FoI_GDR} are sensitive to noise and high-frequency components in the images $I_0,\cdots,I_{N-1}$.
Lung parenchyma has high-frequency texture due to small airways and vessels.  As such, the performance of Algorithm~\ref{al:FoI_GDR} for registering lung images is reduced.
The FoT GDR algorithm (Algorithm~\ref{al:FoT_GDR}) computes the flow of images $I(t)$ in step 4 and the costate $\lambda(t)$ in step 8 using the smooth transformations $\phi_t$.  As a result, Algorithm~\ref{al:FoT_GDR} is not adversely affected by image noise and high-frequency image texture.

\subsection{Generation of Artifact Masks}
\label{sec:mask}
The field of view of a CT scanner does not cover the entire height of the lung, and thus a 4DCT of the lung is acquired by taking stacks of transverse slices over multiple breathing cycles. 
Artifacts often appear at stack boundaries due to variations in breathing, e.g., a deep breath followed by a shallow breath or if the patient is not breathing fast enough, and there is no data for a stack for a particular phase (i.e., a missing phase stack). 
The CT reconstruction algorithm does its best to interpolate these discontinuities or missing data, but this results in the types of artifacts that we are trying to remove. 
The artifact masks that we use in this work remove whole stacks from one of the phases. 
Thus, they appear as rectangles in a coronal slice of the lung.  
The stacks we remove do not necessarily correspond to the imaging stacks, i.e., the stacks that we remove can straddle the imaging stacks where the artifact appears.
	
 We implemented a 2D U-net convolutional neural network (CNN) based on~\cite{ronneberger2015u} to automatically detect motion artifacts in CT images and create artifact masks.
Figure~\ref{fig:unet} in \ref{sec:u-net} shows the CNN  architecture we used.
The input to the network is a $256\times256$ coronal slice from the 3D CT volume.  
The output of the network is a $256\times256$ probability map indicting locations of artifacts in the input image, where 0 corresponds to artifact-free regions and 1 corresponds to artifact regions.
During the training, we used the balanced cross entropy loss~\cite{bhattacharya2020corrsignet}
$L(y_{true}, y_{pred}) = -\sum_{i = 1}^{N} \big( \beta y_{true}^i \log y_{pred}^i + (1-\beta) (1 - y_{true}^i) \log (1 - y_{pred}^i) \big)$, where $\beta = 1- \frac{(\sum_{i = 1}^{N} y_{true}^i) + 0.001}{N + 0.001}$ since the artifact regions only account for approximately 0.5\% of CT pixels. 

\begin{figure}[!hbt]
	\centering
	\includegraphics[width=1.0\linewidth]{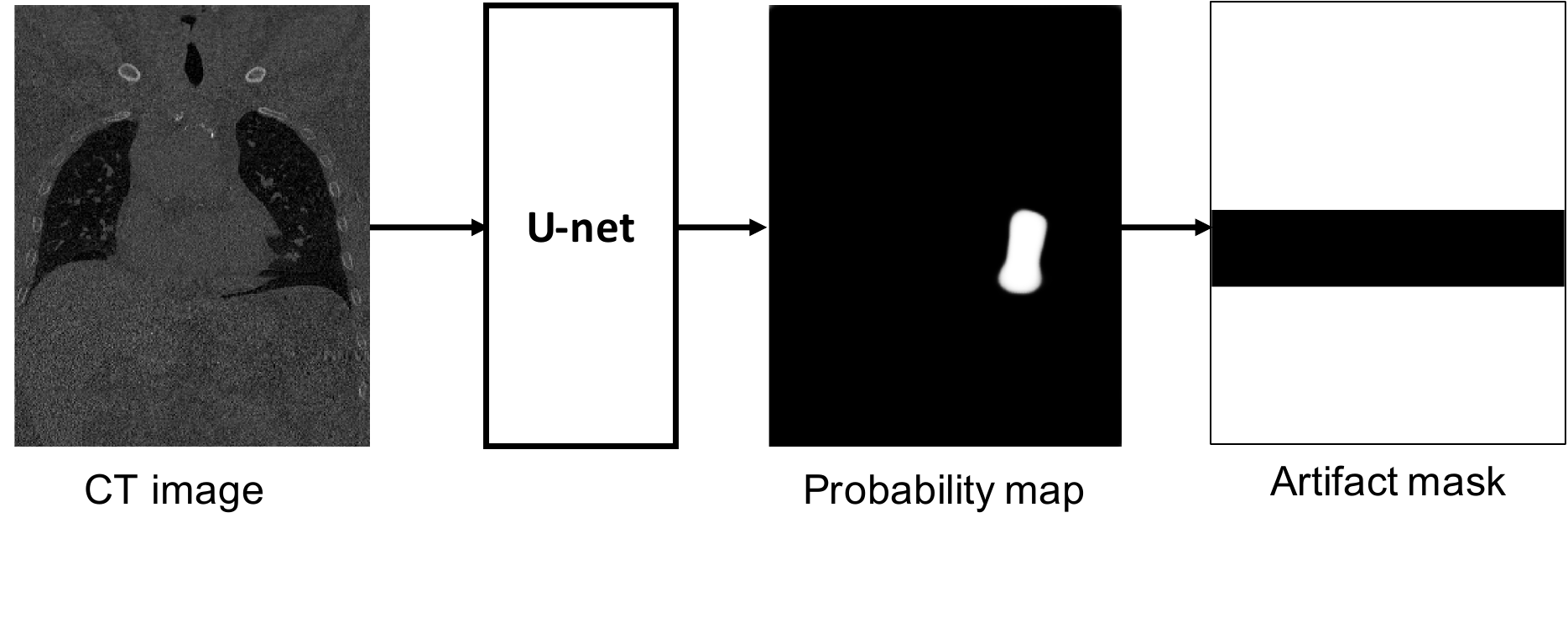}
	\caption{Pipeline for artifact location detection and artifact mask generation.
		\label{fig:pipeline}}
\end{figure}

Figure~\ref{fig:pipeline} shows the pipeline used to automatically detect artifact locations and generate artifact masks for the GDR algorithm.
A coronal CT slice is fed into the U-net model to create an artifact probability map.
The probability map is then thresholded using a threshold of 0.95.
Based on the processed probability map, an artifact mask is created to cover the entire stack containing the local artifact.

\begin{figure}[!hbt]
	\centering
	\includegraphics[width=1.0\linewidth]{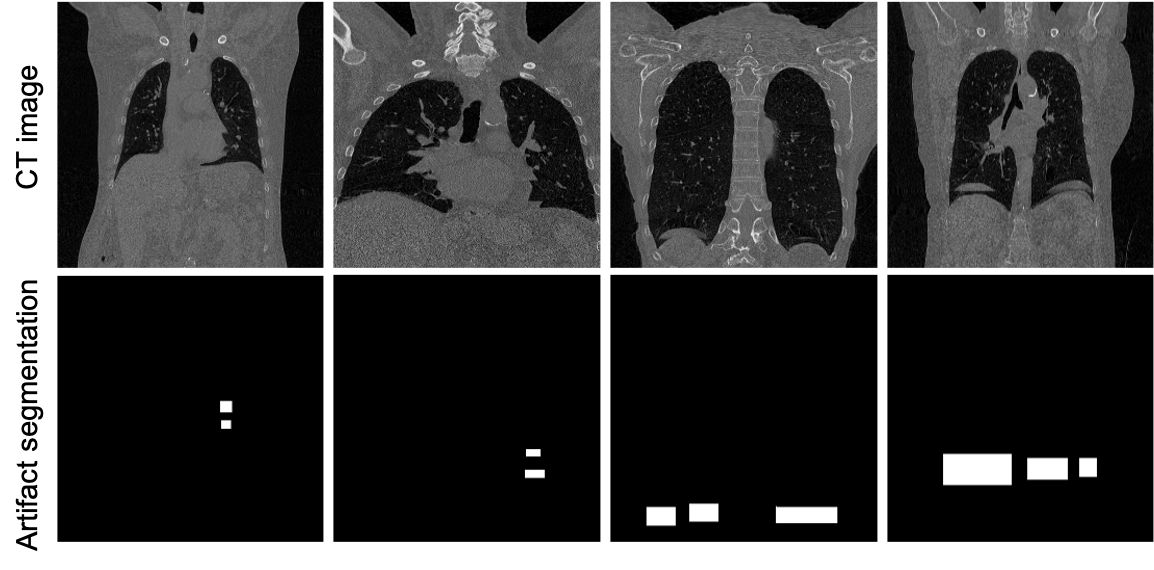}
	\caption{An example of CT images and the corresponding artifact segmentations used to train the artifact detection network.
		\label{fig:unet_training_data}}
\end{figure}

The artifact detection network was trained using 2084 2D coronal slices from 20 CT volumes.
For each 3D CT volume, artifact locations were manually labeled in each coronal slice.
The training dataset consists of CT slices with different amounts and types of artifacts (see Fig.~\ref{fig:unet_training_data} for a few examples).
During the training, we used a learning rate of 0.0003, a batch size of 32, and the Adam optimizer~\cite{kingma2014adam}.

\section{Experiments and Results}
\label{sec:results}
The performance of the GDR algorithm was evaluated using synthesized 2D time-series CT images and treatment planning 4DCT with motion artifacts.
The GDR performance was compared to the state-of-the-art geodesic intensity regression (GIR) algorithm~\cite{NiethammerRegression2011}, which does not modify CT intensities during image regression and does not exclude artifact data from regression.

\subsection{Accuracy of the Artifact Detection Neural Network}
\begin{figure}[!hbt]
	\centering
	\includegraphics[width=1.0\linewidth]{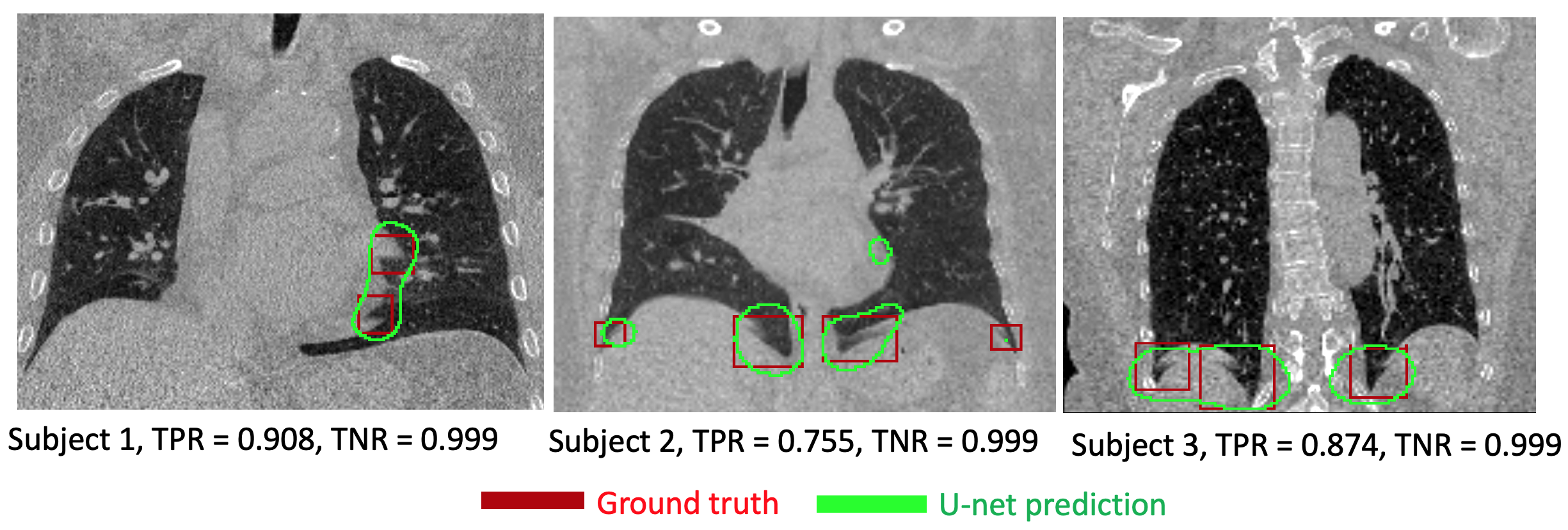}
	\caption{Artifact regions on three different CT volumes predicted by the U-net network. 
		\label{fig:unet_predictions}}
\end{figure}

We used 311 2D coronal slices from three 3D CT volumes for the evaluation of the artifact detection network.
Figure~\ref{fig:unet_predictions} shows a representative result from each CT volume, which demonstrates the accuracy of the artifact detection network.
The average true positive rate (TPR) and the average true negative rate (TNR) of our artifact detection network are 0.846 and 0.999, compared to 0.70 and 0.97 from previous work~\cite{bouilhol2014motion}.

\subsection{2D CT Time-series Simulation}
\label{sec:ct_time_series}

Simulated 2D CT ``artifact-free" time-series were constructed using the following procedure.
We selected ``artifact-free" treatment planning 4DCT scans from 30 human subjects to generate 30 ``artifact-free" CT time-series.
For each 4DCT scan, we selected a coronal slice from the 0EX and 100IN phases of the 4DCT.
Pairwise registration was performed between the 0EX image and the 100IN image using the GDR Algorithm~\ref{al:FoT_GDR}.
The resulting displacement field $u$ was scaled by factors of 0.2, 0.4, 0.6, 0.8, and 1.0 to deform the 100IN image into 80IN, 60IN, 40IN, 20IN, and 0EX breathing phase images, corresponding to time points $t \in \{0.2, 0.4, 0.6, 0.8, 1.0 \}$.
This procedure produced a simulated CT time-series consisting of six breathing phase images with known spatial correspondences.

The intensity of the simulated deformed CT images were adjusted using 
the conservation of mass equation $\frac{\text{CT}(t) + 1000}{1055} = |D\phi_t|\frac{\text{CT} + 1000}{1055}$, where $|D\phi_t|$ is the Jacobian determinant of the transformation $\phi_t = Id + t\cdot u$ .
This equation relates the Hounsfield units of the image CT$(t)$ at time $t$, $0 \leq t \leq 1$ to the Hounsfield units of the original CT image (100IN CT at $t = 1$) by tissue volume scaling.
Solving this equation for $\text{CT}(t) = |D\phi_t|(\text{CT} + 1000) - 1000$ provides a method to properly simulate the intensity of deformed CT images.

\subsection{2D Data Dropout Experiment}
\label{sec:2d_dropout}

\begin{figure}[!htb]
	\centering
	\begin{subfigure}[t]{0.115\textwidth}
		\centering
		\includegraphics[height=0.8in]{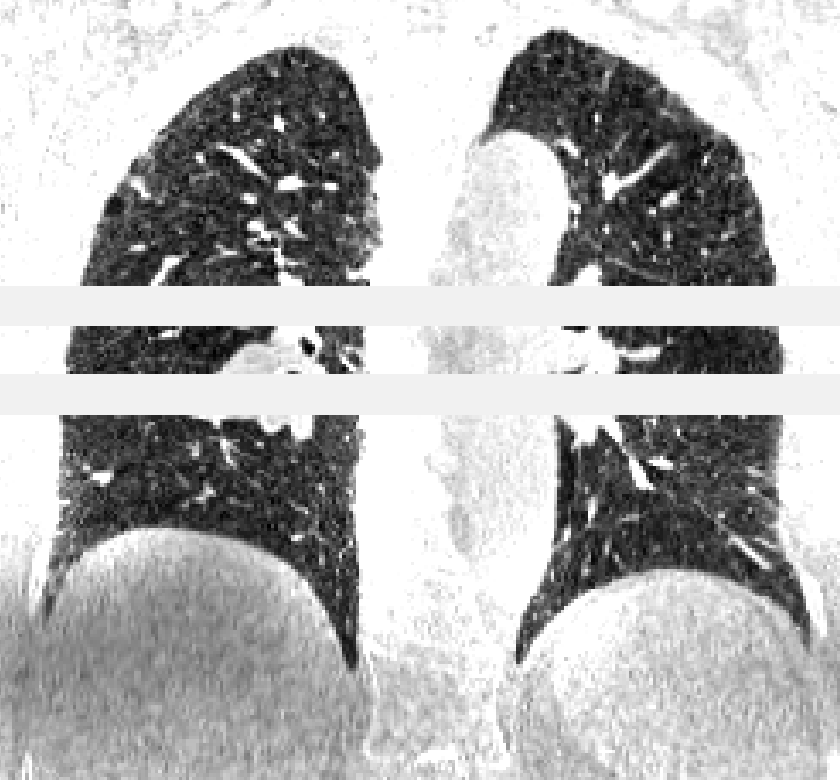}
		\caption{0EX}
	\end{subfigure}
	\hspace{0.025in}
	\begin{subfigure}[t]{0.115\textwidth}
		\centering
		\includegraphics[height=0.8in]{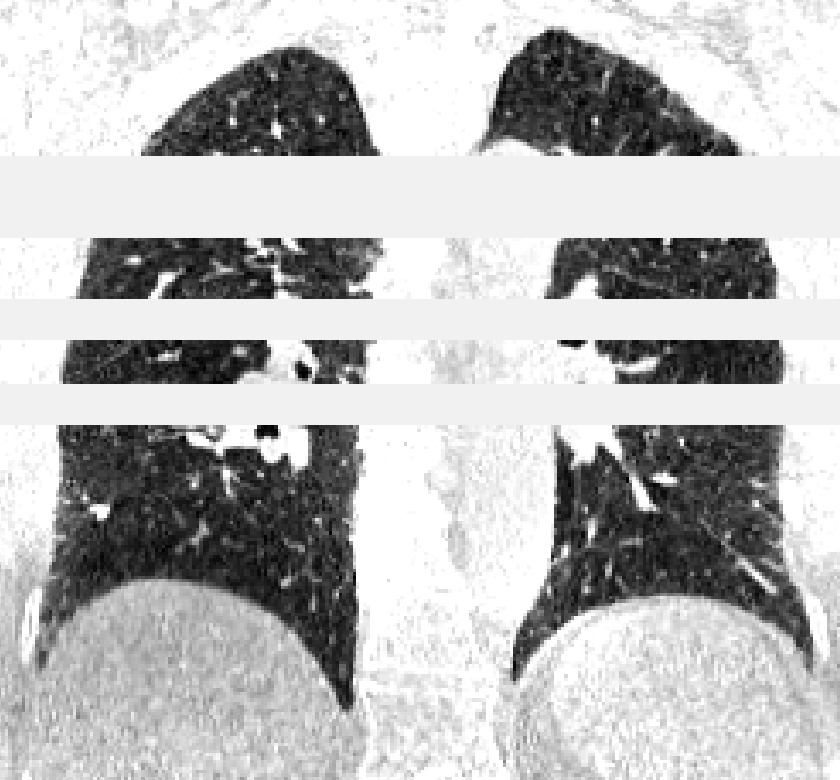}
		\caption{40IN}
	\end{subfigure}
	\hspace{0.025in}
	\begin{subfigure}[t]{0.115\textwidth}
		\centering
		\includegraphics[height=0.8in]{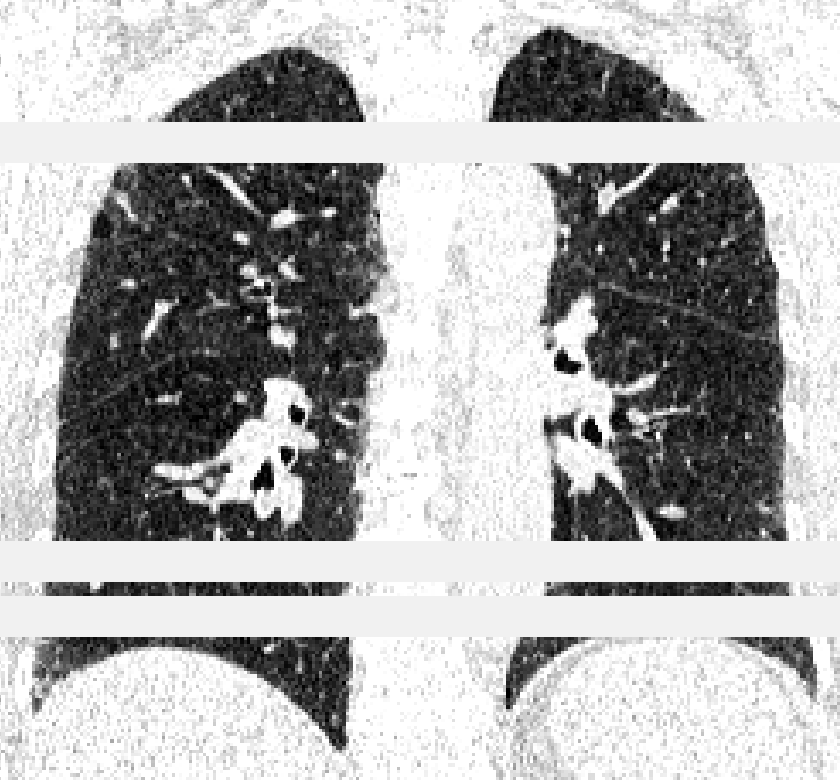}
		\caption{100IN}
	\end{subfigure}
	\caption{An example of 20\% data dropout of a CT time-series. 
		Data in the slabs were excluded from image regression.
		The CT time-series consists of six phases. 
		Only 0EX, 40IN, and 100IN CT images are shown due to space consideration.
		\label{fig:dropout_20Percent}}
\end{figure}

The artifact-free CT time-series data described in Section~\ref{sec:ct_time_series} were used to study the sensitivity of the GDR algorithm to masking out artifact regions.
The idea of this experiment is to pretend that there was an artifact in the masked-out region and to measure how well the algorithm estimated the true Jacobian determinant values.
The locations of 12mm-thick masks were randomly distributed long the $z$ direction over the six breathing phases for each dropout experiment.
We chose the thickness of each data dropout to be 12mm since it is the mean thickness of artifact stacks near the diaphragm or the heart~\cite{YamamotoTokihiro2008RAoA}.
Figure~\ref{fig:dropout_20Percent} shows an example of 20\% dropout for one simulated CT time-series.

\begin{figure}[!hbt]
	\centering
	\includegraphics[width=0.7\linewidth]{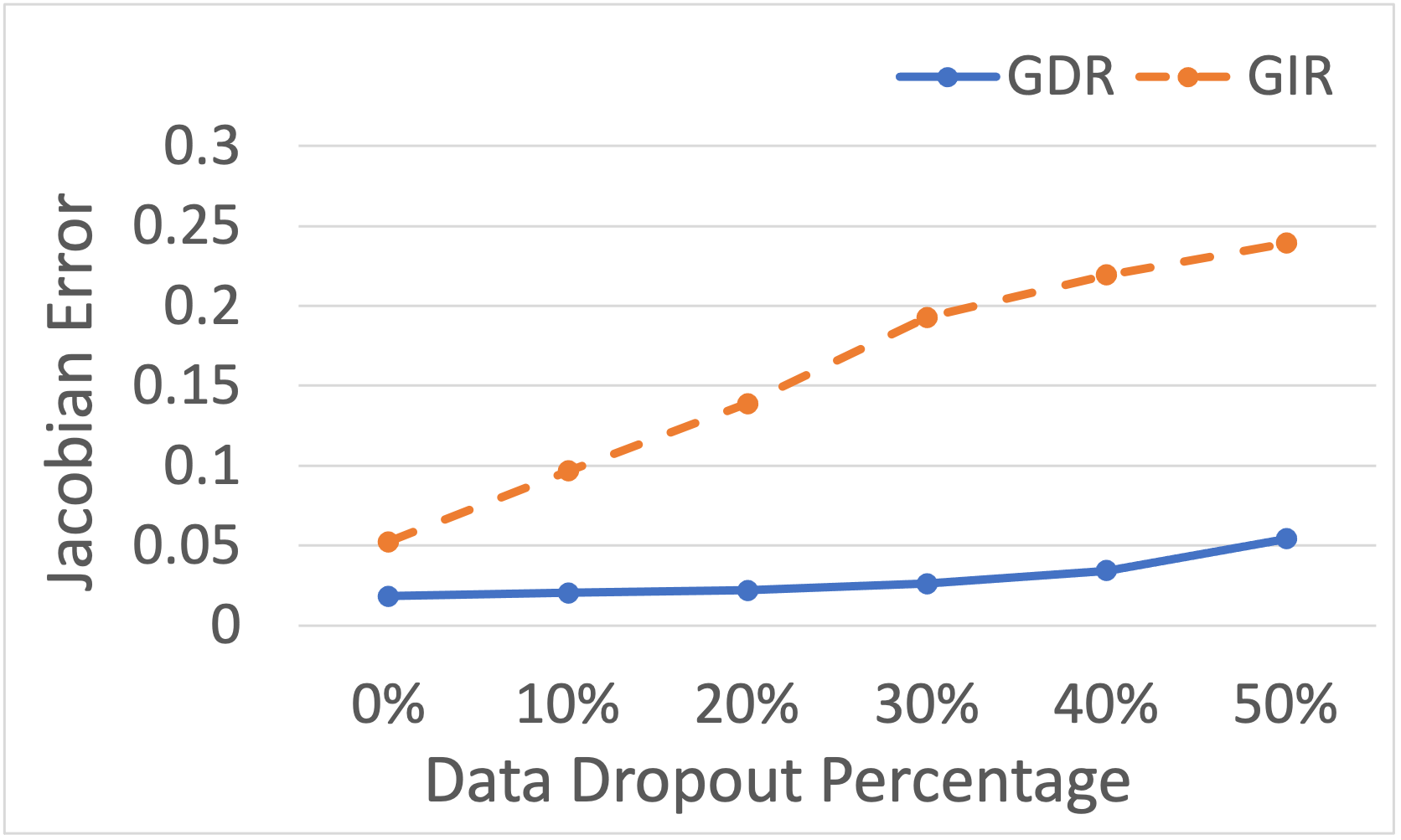}
	\caption{Average Jacobian errors of GDR and GIR regressions of 2D CT time-series with data dropout.
		\label{fig:dropout_sensitivity}}
\end{figure}

Figure~\ref{fig:dropout_sensitivity} compares the mean Jacobian errors of the GDR and GIR regression algorithms at different data dropout levels for this experiment.
The Jacobian error was defined as the average absolute difference between the true Jacobian image and the estimated Jacobian image. 
The true Jacobian images were computed from the transformations used to construct the 30 CT time-series data sets.
For each dropout level, ten different dropout experiments were performed for each simulated time series, i.e., $5\times10\times30 = 1500$ GDR and GIR regressions were performed to generate Fig~\ref{fig:dropout_sensitivity}.
Both the GDR and GIR algorithms used 10 time points between two adjacent phases and multiple resolutions, with kernel sizes ($\sigma$) of 60, 30, 15 mm, downsample factors of 4, 2, 1, weighting factors ($\gamma$) of 0.1, 0.1, 0.1.

Figure~\ref{fig:dropout_sensitivity} shows that the GDR algorithm is less sensitive to data dropout than the GIR algorithm.
Notice that data dropout had little effect on the performance of the GDR algorithm until about 30\% dropout.
Conversely, the performance of the GIR algorithm started declining at 10\% data dropout.
The robustness of our GDR algorithm indicates that the artifact masks do not need to be accurate and the users only need to have a rough estimate of the locations of artifacts when applying GDR to correct 4DCT motion artifacts.

\subsection{2D Duplication Artifact Experiment}
\label{sec:2D_duplicate}

\begin{figure}[!hbt]
	\centering
	\includegraphics[width=0.8\linewidth]{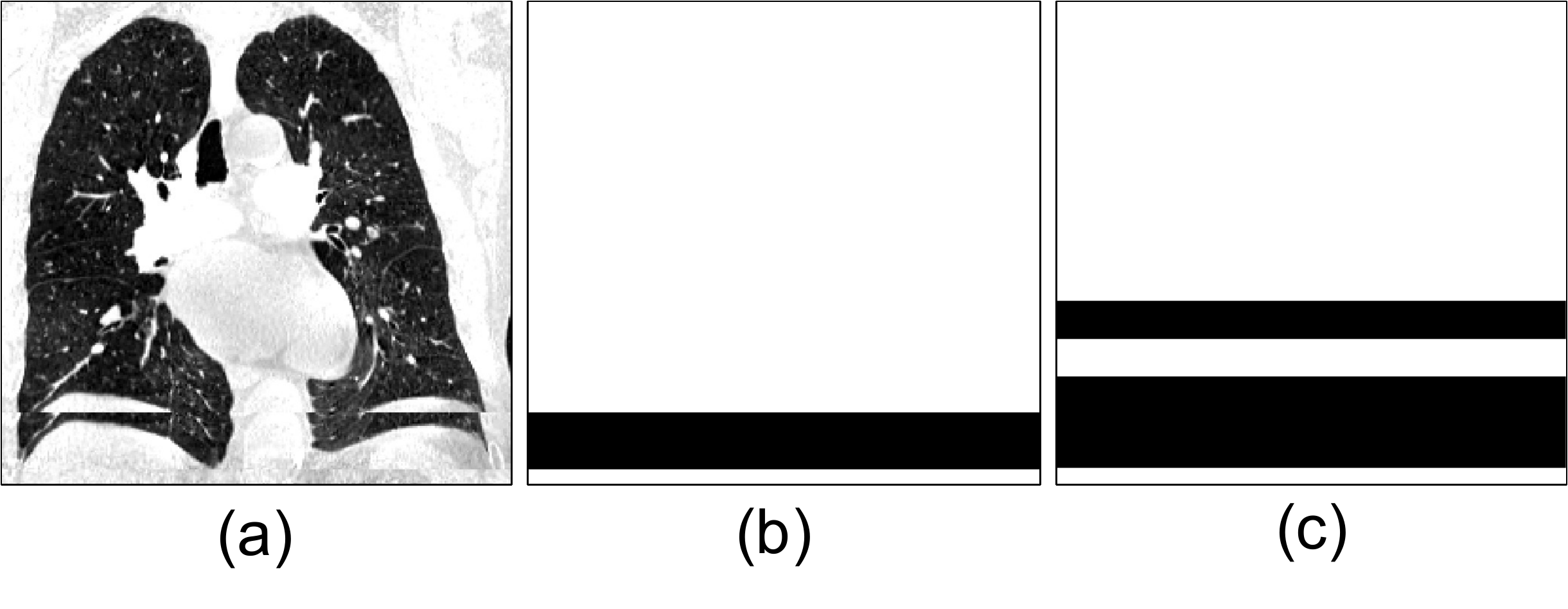}
	\caption{
		An example of choosing the artifact mask for duplication artifact.
		(a) CT with a duplication artifact near the diaphragm; (b) true artifact mask;  (c) artifact mask predicted by U-net.
		\label{fig:2d_IPF004_duplicate}}
\end{figure}

Duplication of the diaphragm is a common motion artifact (see Figs.~\ref{fig:intro_common_artifacts} and \ref{fig:intro_duplicate_artifact}).
To study how well the GDR algorithm removes duplication artifacts, we simulated diaphragm duplication artifacts in one of the six-phases of the 2D CT time-series data sets created in Section~\ref{sec:ct_time_series}.
Figure~\ref{fig:2d_IPF004_duplicate} shows a simulated duplication error  created in the phase 0EX image of one of the six-phase time series. 

\begin{figure}[!htb]
	\centering
	\includegraphics[width=1.0\linewidth]{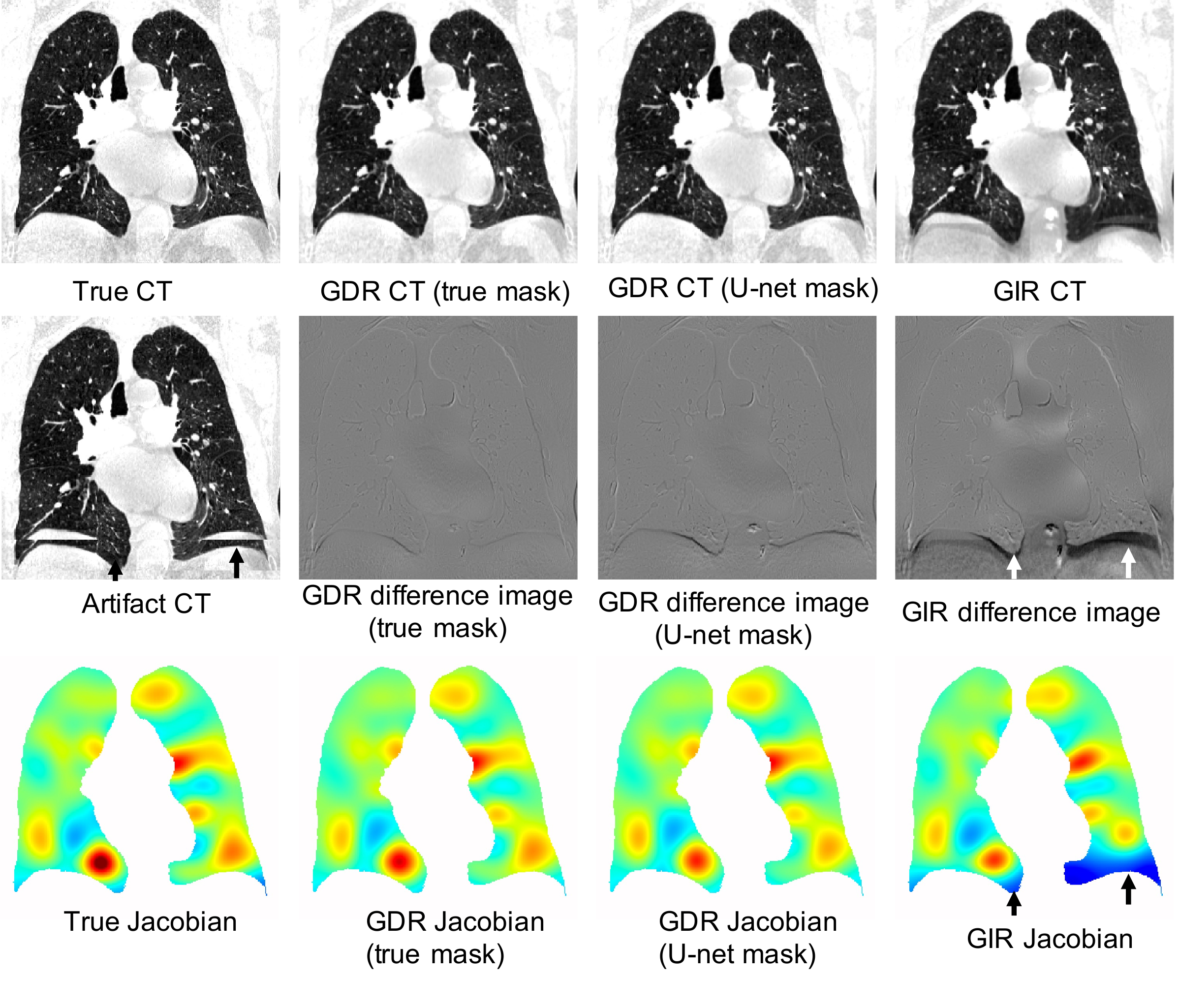}
	\caption{GDR and GIR regression results of a typical subject with simulated duplicate artifact in the 0EX CT of a 6-phase time series. We use the same parameters as in Sec.~\ref{sec:2d_dropout}.
		\label{fig:2D_duplicate_artifact}}
\end{figure}

Figure \ref{fig:2D_duplicate_artifact} shows the regression results produced by the GDR and GIR algorithms for the data set shown in Figure \ref{fig:2d_IPF004_duplicate}.
Panel (b) and panel (c) of Fig.~\ref{fig:2d_IPF004_duplicate} are the true and U-net predicted binary artifact masks used in our GDR regressions.
The true artifact mask denotes the exact location of the simulated duplication artifact and the U-net predicted artifact mask is an estimate of the ground truth.
No artifact mask was used in the GIR algorithm.
We used the same image regressions parameters as in Section~\ref{sec:2d_dropout}.
The difference images in Fig.~\ref{fig:2D_duplicate_artifact} 
show that the GDR algorithm achieved a more accurate estimate of the 0EX CT image than the GIR algorithm, especially near the artifact region (white arrows).
The Jacobian images show that the GDR algorithm produced an accurate Jacobian image close to the true Jacobian, whereas the GIR algorithm produced a Jacobian image with a significant artifact near the diaphragm.
The similarity between GDR results produced by using two different artifact masks also demonstrate the robustness of our GDR algorithm to the choice of artifact mask.
These results illustrate the advantage of using artifact masks to exclude artifact data from regression.

We repeated the experiment shown in Fig.~\ref{fig:2D_duplicate_artifact} 180 times using the following procedure.
For each of the 30 artifact-free CT time-series generated in Section~\ref{sec:ct_time_series}, we generated six CT time-series with a diaphragm duplication artifact of 30 mm in one of the six breathing phases.
We applied the GDR (with true artifact masks) and GIR algorithms to these 180 motion artifact simulated CT time-series.
The average Jacobian errors for the GDR and GIR regressions in the artifact regions were 0.03 and 0.12, respectively.
A one-tailed t-test shows that the GDR algorithm produced a significantly more accurate (p-value $<$ 0.001) Jacobian image than the GIR algorithm.
GDR regressions using artifact masks predicted by the U-net resulted in an average Jacobian error of 0.037. 
There was no significant difference (p = 0.31) between GDR results generated by the true and U-net predicted artifact masks. 

\subsection{4DCT Motion Artifact Experiment}

\begin{figure}[!hbt]
	\centering
	\includegraphics[width=0.8\linewidth]{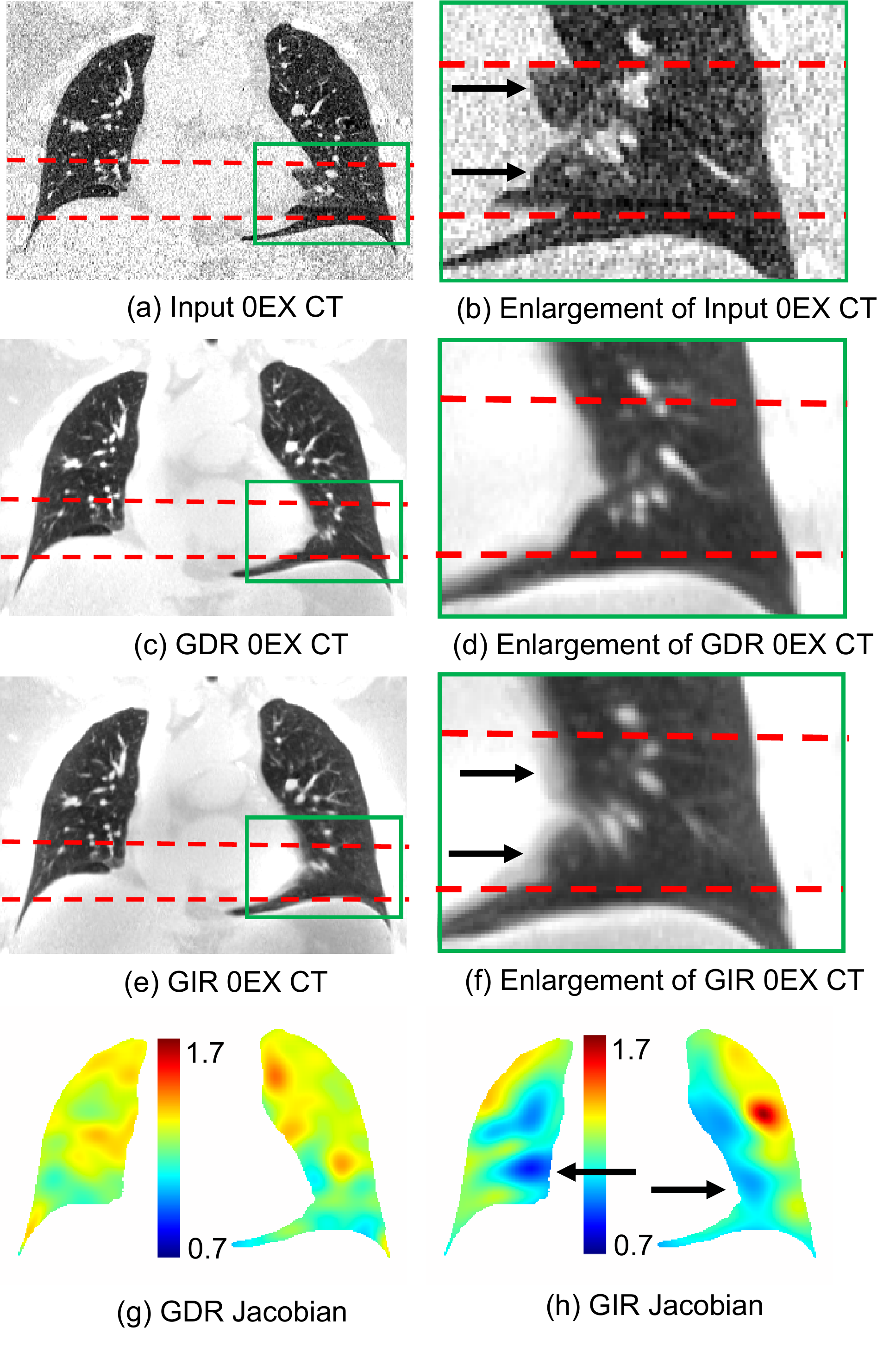}
	\caption{ GDR and GIR regression results of a 4DCT scan with misalignment artifacts. 
		Artifact data between the dashed lines were excluded from the GDR regression. 
		The Jacobian images were computed in the coordinate system of the 0EX CT.
		The arrows are pointing at artifact regions.
		\label{fig:IPF003_misalignment}}
\end{figure}

Figure~\ref{fig:IPF003_misalignment} shows the results of applying the GDR and GIR algorithms to a treatment planning 4DCT scan (See Section \ref{sec:4DCTdata}) with misalignment artifacts.
Both algorithms aligned the 10 phase volumes using five time points between two adjacent phases and multiple resolutions, with kernel sizes ($\sigma$) of 48, 24, 12 mm, downsample factors of 4, 2, 1, weighting factors ($\gamma$) of 0.05, 0.075, 0.1.
Artifact data between the dashed lines were excluded from the GDR regression.
Figure~\ref{fig:IPF003_misalignment}(c-d) show that the GDR algorithm successfully corrected the misalignment artifact in the 0EX CT and produced a sharp 0EX CT image.
Figure~\ref{fig:IPF003_misalignment}(e-f) show that the GIR algorithm was also able to mitigate the misalignment artifact but is blurry (arrows) because it averaged data with artifacts.
The results also show that both the GDR and GIR regressions improved the signal-to-noise ratio (SNR) of the CT images.
Figure~\ref{fig:IPF003_misalignment}(g-h) show that even though the GDR and GIR algorithms gave similar results with respect to intensity, the Jacobian determinant images are very different in the regions near the artifacts.
Note that Fig.~\ref{fig:IPF003_misalignment} only shows the artifact masks that were used in the 0EX phase, but artifacts were masked in other phases as well. 
This explains why the Jacobian determinant is different in areas of the 0EX phase besides the region that was masked out in the 0EX phase.

\begin{figure}[!hbt]
	\centering
	\includegraphics[width=2.2in]{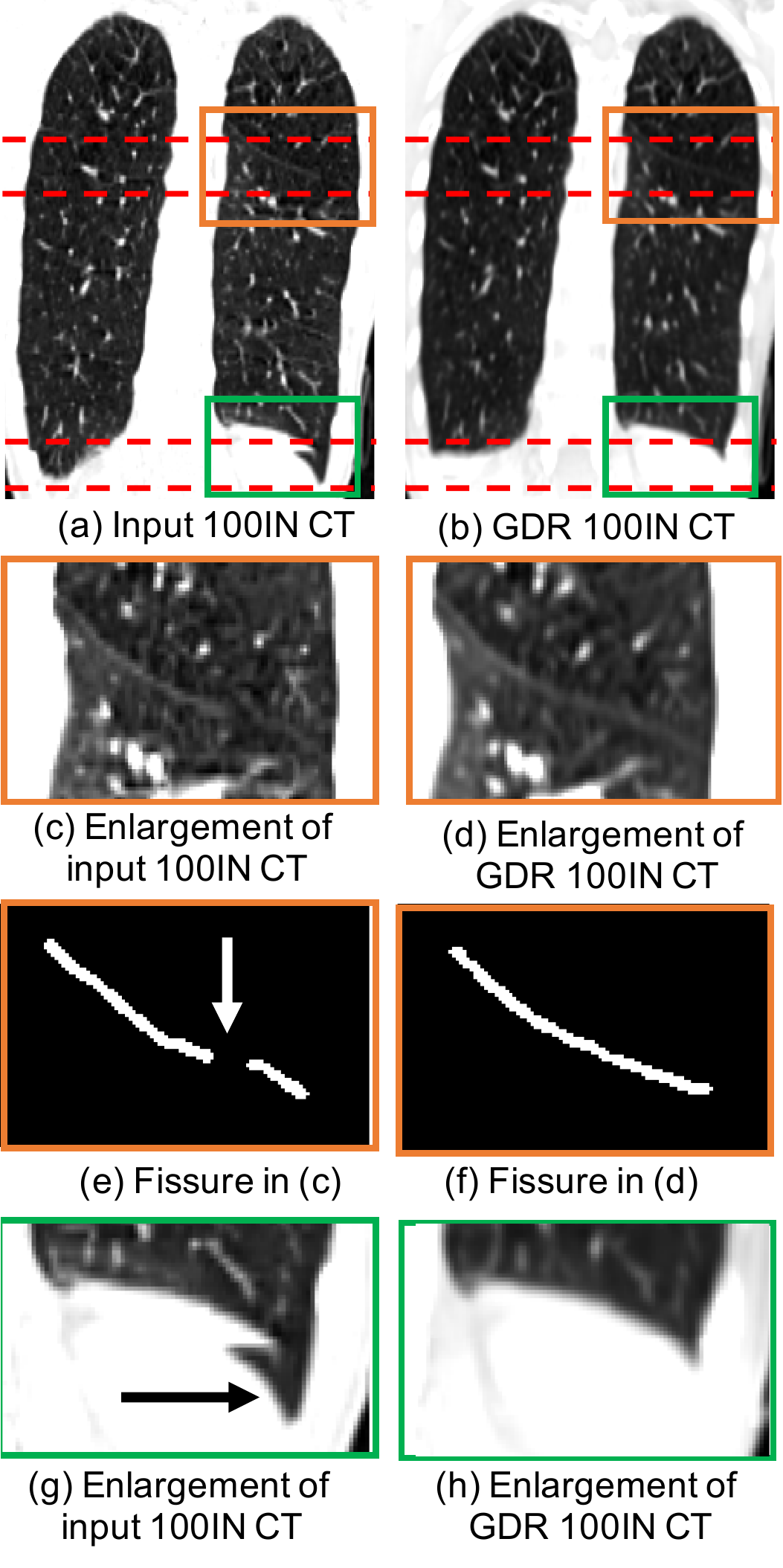}
	\caption{ GDR and GIR regression results a 4DCT scan with duplication artifacts.
		Artifact data between the dashed lines were excluded from the GDR regression.
		The rectangular boxes correspond to regions enlarged for visualization in the second, third, and forth rows.
		\label{fig:IPF003_duplicate}}
\end{figure}

Figure~\ref{fig:IPF003_duplicate} shows the result of applying GDR to a 4DCT scan with discontinuous lung fissure and a duplication artifact near the diaphragm.
Artifact data between the dashed lines were excluded from the GDR regression by using an artifact mask covering the artifact regions.
The second row of Fig.~\ref{fig:IPF003_duplicate} shows the enlarged CT images in the orange rectangular boxes from the first row.
The third row shows the manually labeled fissures before and after applying the GDR algorithm.
Notice that the GDR algorithm successfully fixed the discontinuous fissure in the input 100IN CT.
The last row of Fig.~\ref{fig:IPF003_duplicate} shows that the GDR algorithm removed the duplication artifact in the 100IN CT.

\subsection{Robustness of the Jacobian Image to Missing Data}

\begin{figure}[!hbt]
	\centering
	\includegraphics[width=0.7\linewidth]{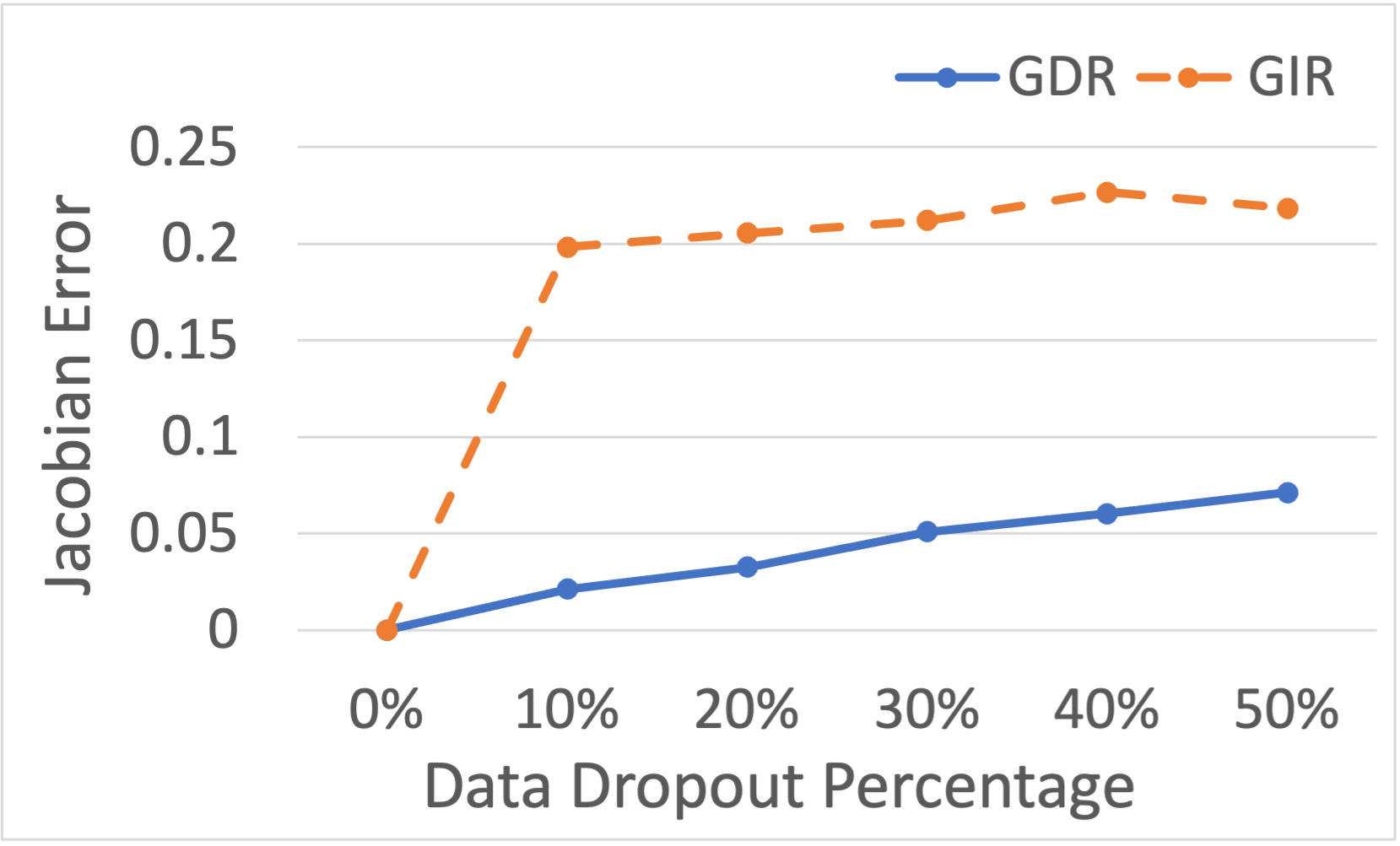}
	\caption{Jacobian change of GIR and GDR with respect to the percentage of data dropout, respectively.
		Ten subjects were used to generate this graph.
		\label{fig:jacobian_robustness}}
\end{figure}

Without knowing the truth, it is not possible to compare the accuracy of the Jacobian images produced by GDR and GIR regressions of treatment planning 4DCT.
Therefore, we studied the robustness of GDR and GIR Jacobian images to missing data, i.e., we studied how robust the regression was with respect to masking out regions of the lung.
In this experiment, we selected 4DCT scans of ten patients with minimal artifacts.
These data sets were regressed using GDR and GIR to set the baseline performance for GDR and GIR, respectively.
Then for each subject, we generated ten data sets with random dropout at each of 10\%, 20\%, 30\%, 40\%, and 50\% dropout levels similar to our experiment in Section~\ref{sec:2d_dropout} for a total of 500 simulated 4DCT data sets.
We then performed GDR and GIR regression on each of these 4DCT data sets.
The GDR method used the true masked-out locations while the GIR did not use any artifact masks.
Figure~\ref{fig:jacobian_robustness} shows the errors of GDR and GIR in estimating the Jacobian images.
Notice that the error in the GDR method gradually increases with the amount of missing data but is still relatively small even at 50\% dropout.
The GIR method has worse performance and is more sensitive to missing data than the GDR algorithm.
The robustness of the GDR algorithm to missing data allows the artifact masks to fully cover artifact regions by using larger artifact masks, which makes the process  of labeling artifact masks more efficient.

\subsection{Image Regression Accuracy Evaluated by Mean Landmark Error}
\begin{figure*}[!hbt]
	\centering
      \includegraphics[width=0.7\linewidth]{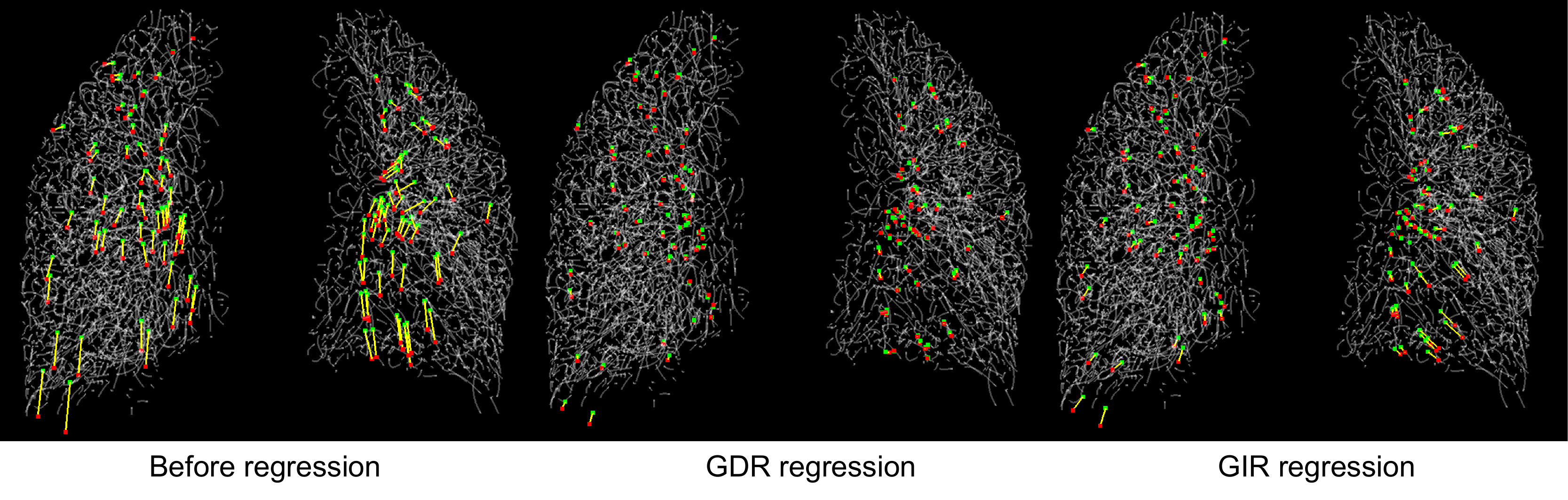}
	\caption{Comparison of landmark matching accuracy of the GIR and GDR regressions for a representative subject. The red points are landmarks in the 100IN phase and the green points are the corresponding landmarks in the 0EX phase. Yellow lines show the landmark errors between corresponding landmarks.
		\label{fig:lmk_err}}
\end{figure*}

Image registration accuracy was evaluated by computing 
the mean landmark error (MLE).
Seven 4DCT scans with minimal motion artifacts were landmarked. 
The IX software~\cite{murphy2008} was used to automatically select 100-125 evenly-distributed landmarks in the 0EX phase, and the corresponding landmarks in the 100IN phase were manually labeled using a landmarking tool~\cite{ZhaoBowen2016Tpdi}.
For each of the seven 4DCT scans, we simulated 5\% data dropout.
Figure~\ref{fig:lmk_err} compares the spatial distributions of landmarks in the 0EX and 100IN phases before and after the GIR and GDR regressions of a representative artifact 4DCT scan. 
The average MLE was reduced from 7.3 mm to 1.25 mm and 2.4 mm by the GDR and GIR regressions, respectively.
A one-sided t-test shows that the GDR regression achieved a significantly ($p<0.01$) smaller MLE than the GIR regression, demonstrating promising performance of the GDR regression for aligning of 4DCT scans with artifacts.
Note that introducing a 5\% dropout artifact to the 4DCT scans increased the average MLE from 1.16 mm to 1.25mm but this 7.8\% change was not significant (p = 0. 62). 

\section{Discussion}
\subsection{Clinical Impact}
Radiation therapy (RT) is commonly used to treat lung cancer patients~\cite{DelaneyGeoff2003Amfd}.
During RT, irradiation of healthy lung tissues will cause damage to the lung.
Since high-function lung tissues are more susceptible to radiation dose than low-function lung tissues~\cite{Patton2018}, minimizing radiation dose to high-function lung can potentially reduce radiation-induced injury of the lung.
This type of RT is called functional avoidance RT~\cite{SivaShankar2016GMP4,ireland2016functional,Yaremko2007562,huang2013imrt,VinogradskiyYevgeniy2016RLFP,yamamoto2016first}.
The proposed GDR algorithm may improve functional avoidance RT for lung cancer patients by improving the accuracy of computed Jacobian images~\cite{reinhardt2008a,kipritidis2019vampire,wei2018TIA,Shao2019OutOfPhase,wallat2020modeling} derived from image registration of 4DCT. 
The quality of the resulting Jacobian images is sensitive to the choice of image registration approach and motion artifacts in 4DCT~\cite{castillo2017numerical,Shao2018Sensitivity}.
We have shown that the proposed GDR algorithm can remove major artifacts in both 4DCT and 4DCT-derived Jacobian, and also improve the signal-to-noise ratio of CT images.
Therefore, the GDR algorithm can provide a more accurate Jacobian map to improve functional avoidance radiation therapy for lung cancer patients.
In addition, using the GDR algorithm to correct 4DCT artifacts may improve the clinical outcome in stereotactic body radiation therapy  of lung and liver metastases~\cite{sentker20204d}.

\subsection{Sharpness of Estimated Template Image}
The GDR approach estimates a template image and its flow over
all breathing phases.
The template image is computed by pulling back all the phases to a common coordinate system using estimated correspondences and averaging all ten deformed 3D phase images. 
The template image may look blurry if the estimated correspondences are not accurate due to averaging.
The estimated template image displayed in Fig.~\ref{fig:IPF003_misalignment}, panel (c) and Fig.~\ref{fig:IPF003_duplicate}, panel (b) show that there is only negligible blurring in the template image, demonstrating that our GDR algorithm has achieved very accurate correspondences.
The sharper the template image is, the better the image regression.

Image sharpness can be quantified by the local variations in image intensity and can be measured using the maximum local variation (MLV) measure~\cite{bahrami2014fast}.
The MLV of a voxel $I_{i,j,k}$ (a voxel at location $(i,j,k)$ of the image $I$) is defined as the maximum variation between the intensity of $I_{i,j,k}$ with respect to its 26-neighbor voxels given by
\begin{equation}
    MLV(i,j,k) = max(|I_{i,j,k} - I_{x,y,z}|).
    \label{eq:MLV}
\end{equation}
where $ i-1 \leq x \leq i+1, j - 1\leq y \leq j+1, k - 1 \leq z  \leq k+1$.
This definition implies that the larger the MLV, the sharper the image is at the location $(i,j,k)$. 

\begin{figure}[!hbt]
	\centering
	\includegraphics[width=0.8\linewidth]{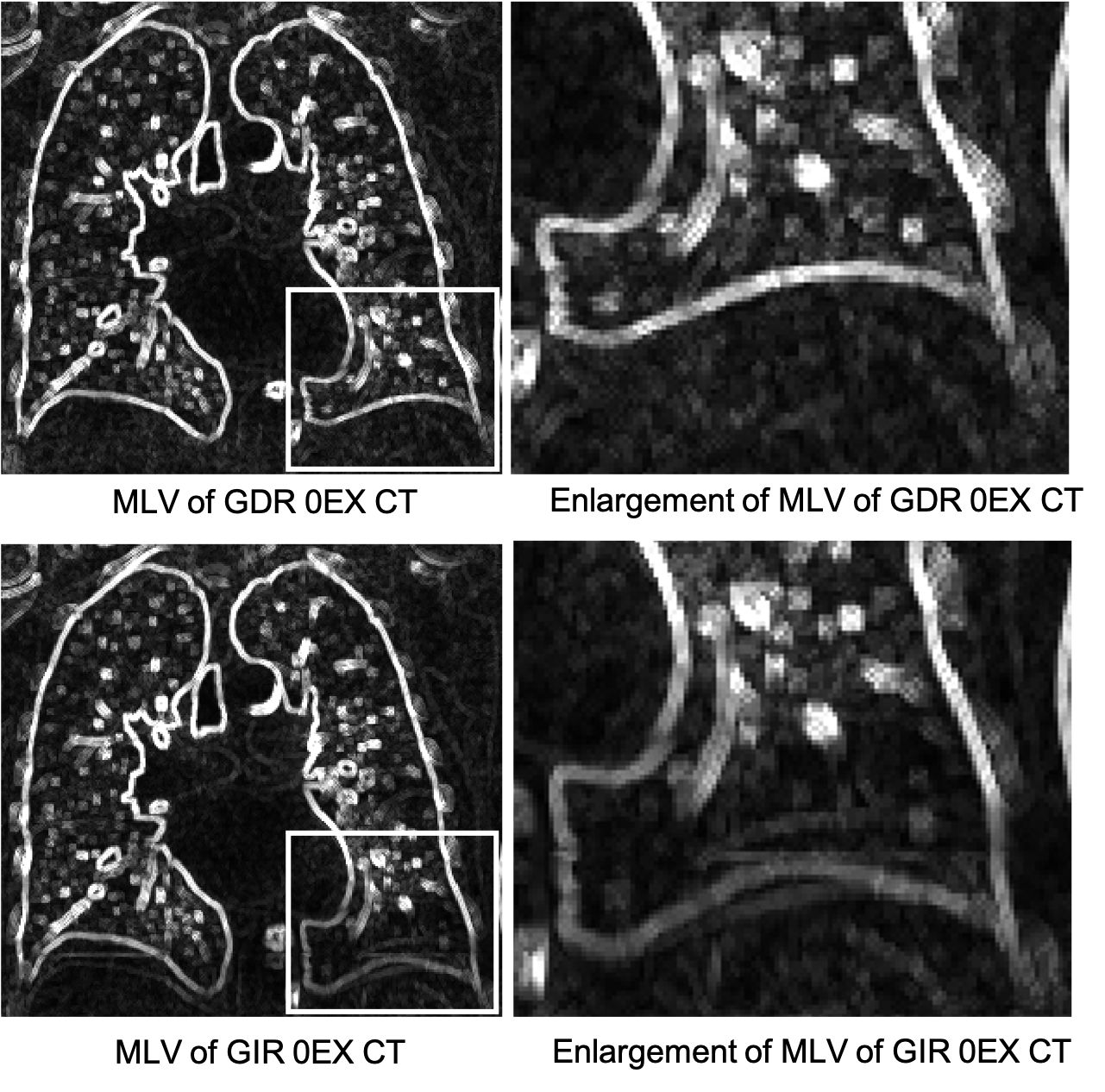}
	\caption{
Maximum local variation (MLE) measures of the template images produced by GDR and GIR regressions on a CT time series with simulated duplication artifacts.
The right column is an enlargement of the artifact region for better comparison.
The brighter the MLE, the sharper the template images are. 
		\label{fig:MLV}}
\end{figure}

Figure~\ref{fig:MLV} shows the MLV images of the GDR and GIR template images in Fig.~\ref{fig:2D_duplicate_artifact}.
This figure shows that the GDR image is sharper than the GIR image in the artifact regions.
Using 20 simulated 4DCT scans of 10 subjects in the 3D duplication artifact experiment, the mean MLVs of the GDR and GIR template images in  the lung regions were 17.5 and 14.8, respectively. 
The template images generated by the GDR regression were significantly ($p<0.001$, one-tailed t-test) sharper than the template images generated by the GIR regression.

\subsection{Implementation Considerations}

We implemented the GDR algorithm on a single high-memory Argon Phase 1 compute node on the University of Iowa 
High-Performance Argon Cluster (\url{https://hpc.uiowa.edu/}).
The node has two Xeon E5-2680v4 (28 Cores at 2.4GHz) for a total of 56 compute slots, 512GB RAM, 1Gbps Ethernet, 100Gbps Omnipath, and HPL benchmark performance of 766.1 GFlops, and costs approximately $\$$7,500.

We used the limited-memory Broyden–Fletcher–Goldfarb–Shanno (L-BFGS) optimizer~\cite{liu1989limited} to achieve fast convergence. 
At each iteration, L-BFGS computes an estimate of the descent direction using the past $m$ updates of the velocity field $v_t$ and the gradient $\nabla_v E(t)$. 
We implemented the L-BFGS optimizer with 20 time points and $m=3$.
The memory requirements for the GDR algorithm can be computed as follows.
The memory to store a $320 \times 320 \times 320$ 3D CT lung image at $1 mm^3$ resolution represented by single-precision floating-point format is $320 \times 320 \times 320 \times 4$ bytes = 131 MB. 
Each 4DCT scan consists of ten 3D breathing phases and we used two time points between adjacent phases. 
With these parameters, the minimum number of 3D image volumes that needs to reside in memory at once is 10 input CT images, 10 input artifact masks, 20 state images, 20 costate images, 20 Jacobian determinant images, 20 push-forward displacement vector fields, 20 pull-back displacement vector fields, 20 velocity vector fields, and 20 gradient vector fields.
Vector valued quantities such as velocity fields require storage of $x$, $y$ and $z$ components and thus require three 3D volumes at each time point.
The resulting memory requirement to hold this data is  $(10+10+20+20+20+20\times3+20\times3+20\times3+20\times3)\times0.131$ GB $\approx$ 42 GB of computer memory. 
The L-BFGS optimizer requires an additional $3\times(1+1)\times20\times3\times0.131$ GB $\approx$ 47 GB memory to store the location and gradient information from the previous three iterations. 
The total memory requirement for the GDR algorithm is approximately doubled to 180GB when counting temporary image volumes used for computation.

The GDR implementation took approximately 6 hours to run which is comparable to the time required to perform 19 traditional 3D pairwise image registrations. 
Originally, we implemented the algorithm using 50 time points with gradient descent.
This gradient descent approach took approximately 50 hours to compute to achieve the same landmark accuracy.

Due to the large memory requirements, we did not implement the algorithm on a GPU.
The current memory limit for state-of-the-art GPUs (e.g., NVIDIA Quadro RTX 8000 and NVIDIA Tesla V100) is 48 GB memory.
However, implementing the GDR algorithm on the GPU holds the potential to greatly increase the speed of the algorithm.
Recent work has shown that running pairwise diffeomorphic registrations on GPUs yielded over 20x speed-up (Brunn et al. 2021).  
In the future, we plan to investigate implementing and running our GDR regressions on multiple GPUs linked together via NVLink (\url{https://www.nvidia.com/en-us/data-center/nvlink/}).



A simple and straight forward way to speed up and reduce the memory requirements of the GDR algorithm is to reduce the image size. 
Downsampling the input images by a factor of two in all dimensions reduces the running time and memory requirement by a factor of $2^3=8$. 
Another approach to reduce the memory requirements of the GDR algorithm is to use stationary velocity fields instead of time varying velocity fields. 
Using stationary velocity fields may greatly reduce the memory requirements of the algorithm since all intermediate velocity fields could be represented as a single velocity field.
Thus, instead of storing 20 time varying velocity fields, one could store 10 stationary velocity fields, i.e., one velocity field between 3D phase image.
Previous work~\cite{risser2013piecewise,vercauteren2008symmetric} showed that using a slightly less accurate stationary velocity field could reduce the registration time by approximately 5 times. 
The cost of using a stationary velocity fields is that it reduces the degrees of freedom of the algorithm and thus may reduce the registration accuracy to a degree.

\subsection{Accommodating Sliding Motion}
One important consideration of lung image registration is to model sliding along chest wall and fissure boundaries. 
Many methods have been proposed to measure~\cite{amelon2014measure} and accommodate sliding motion in image registration~\cite{risser2013piecewise,papiez2014implicit}. 
The GDR method can accommodate the sliding motion to a degree by the fact that it composes many small smooth transformations to estimate a large deformation.
The composition of many small smooth transformations can represent/model narrow areas of shearing, i.e., sliding along lung boundaries. 
The continuous nature of GDR regression may introduce some artifacts inside the lungs. The degree of the artifact inside the lung is dependent on many factors including the number of intermediate time points, the resolution of the images and the degree of sliding present in the data set.
In the future, we plan to investigate how well the GDR algorithm can accommodate the sliding motion.

\section{Conclusions}

We developed a geodesic density regression (GDR) algorithm to correct artifacts caused by motion present in 4DCT lung images.
GDR estimates the geodesic flow of each point of the lung tissue from one phase of the 4DCT to the next while being robust to motion artifacts.
The GDR algorithm takes into account CT intensity change caused by local tissue density change during breathing.
The GDR algorithm uses binary artifact masks to exclude bad artifact image data from being used in the regression.
The GDR approach is robust to large regions of dropout because of the redundancy in the 4DCT data, i.e., dropping out regions of intensity from one phase can be inferred by using corresponding image regions from past and future phases.
Averaging corresponding phase data across phases allows the GDR algorithm perform well in artifact regions even without using a mask.
In this case, the effect of the artifact region is mitigated by averaging it with good data from the other phases.
Results show that the GDR algorithm is effective for removing both simulated and clinically observed 4DCT motion artifacts.


\appendix

\section*{APPENDIX}
\setcounter{section}{1}

\subsection{Computing Time Derivative of the Density Flow}
\label{sec:derivative_of_density_flow}
The time derivative of the density flow in Eq.~\ref{eq:density_flow} is given by
\label{sec:time_derivative_image_flow}
\begin{equation}
\begin{split}
\dot{I}(t)
& = \frac{d}{dt} \big(|D\phi_t^{-1}|I_T\circ \phi_t^{-1}\big)\\
& = \frac{d}{dt}(|D\phi_t^{-1}|)I_T\circ \phi_t^{-1} + |D\phi_t^{-1}|\frac{d}{dt}(I_T\circ \phi_t^{-1}) \\
& = tr\big(adj(D\phi_t^{-1})\frac{d}{dt}(D\phi_t^{-1}) \big)I_T\circ \phi_t^{-1}\\
&  ~~~~ + |D\phi_t^{-1}|\big(-\nabla(I_T\circ \phi_t^{-1})\cdot v_t\big).
\end{split}
\label{eq:time_derivative_image_flow}
\end{equation}

If $f:\mathbb{R}^3\rightarrow\mathbb{R}$ and $g:\mathbb{R}^3\rightarrow\mathbb{R}$ are smooth functions, then the gradient of $fg$ is given by $\nabla(fg) = f\nabla(g) + \nabla(f)g$. Applying this product rule for gradient to $f = ||D\phi_t^{-1}||$ and $g = I_T\circ \phi_t^{-1}$ we have 
\begin{equation}
\begin{split}
\nabla I(t) & = \nabla (|D\phi_t^{-1}|I_T\circ \phi_t^{-1}) \\
& = |D\phi_t^{-1}|\nabla(I_T\circ \phi_t^{-1}) + \nabla (|D\phi_t^{-1}|) I_T\circ \phi_t^{-1}.
\end{split}
\label{eq:gradient_image_flow}
\end{equation}
Notice that 
\begin{equation}
\nabla (|D\phi_t^{-1}|)_i = \text{vec}(adj^{T}(D\phi_t^{-1}))\cdot \text{vec}(\frac{\partial}{\partial x_i}D\phi_t^{-1}).
\label{eq:gradient_of_determinant}
\end{equation}
Combining Eq.~\ref{eq:gradient_image_flow} and Eq.~\ref{eq:gradient_of_determinant}, $|D\phi_t^{-1}|\big(-\nabla(I_T\circ \phi_t^{-1})\cdot v_t\big)$ becomes
\begin{equation}
-\nabla I(t)\cdot v_t + I_T\circ\phi_t^{-1}\sum_{i=1}^{3}\text{vec}(adj^{T}(D\phi_t^{-1}))\cdot \text{vec}(\frac{\partial}{\partial x_i}D\phi_t^{-1})v_t^{i}.
\label{eq:second_term}
\end{equation}
Since we can exchange time and spatial derivatives, then
\begin{equation}
\begin{split}
\frac{d}{dt}(D\phi_t^{-1}) &= D(\frac{d}{dt}\phi_t^{-1}) \\
&= -D(D\phi_t^{-1}v_t)\\
&= -[\frac{\partial}{\partial x_1}D\phi_t^{-1}v_t,\frac{\partial}{\partial x_2}D\phi_t^{-1}v_t,\frac{\partial}{\partial x_3}D\phi_t^{-1}v_t] \\
& ~~~~- D\phi_t^{-1}Dv_t.
\end{split}
\end{equation}
Then $tr\big(adj(D\phi_t^{-1})\frac{d}{dt}(D\phi_t^{-1}) \big)$ becomes
\begin{equation}
-\sum_{i=1}^{3}\text{vec}(adj^{T}(D\phi_t^{-1}))\cdot \text{vec}(\frac{\partial}{\partial x_i}D\phi_t^{-1})v_t^{i}- |D\phi_t^{-1}|tr(Dv_t).
\label{eq:first_term}
\end{equation}
Combining equations~\ref{eq:time_derivative_image_flow},~\ref{eq:second_term} and~\ref{eq:first_term} gives
\[
\dot{I}(t) = -I(t)\text{div}(v_t) - \triangledown I(t)\cdot v_t = -\text{div}\big(I(t)v_t\big).
\]

\subsection{Formula for Updating the Template Image}
\label{sec:update_template}
The GDR energy function (Eq.~\ref{eq:reg_cost}) can be rewritten as
\[
E = E(\phi_t) + \sum_{i=0}^{N-1} h_i(I_T)
\]
where $E(\phi_t)$ is defined in Eq.~\ref{eq:smothness} and 
$h_i(I_T)$ is given by
\begin{equation}
\begin{split}
h_i(I_T) & = \frac{1}{\gamma^2}||(\phi_{t_i}\cdot I_T - I_i)M_i||_{L_2}^2 \\
& = \frac{1}{\gamma^2}\int_{\mathbb{R}^3}\Big( |D\phi_{t_i}^{-1}|(x)I_T\circ\phi_{t_i}^{-1}(x) - I_i(x)\Big)^2 M_i(x) dx \\
& = \frac{1}{\gamma^2}\int_{\mathbb{R}^3} \Big( |D\phi_{t_i}^{-1}|\circ\phi_{t_i}(y)I_T(y) - I_i\circ\phi_{t_i}(y)\Big)^2 \\
& ~~~~~~~~~~\times M_i\circ\phi_{t_i}(y) |D\phi_{t_i}|(y) dy \\
& = \frac{1}{\gamma^2}\int_{\mathbb{R}^3} \Big( |D\phi_{t_i}^{-1}|\circ\phi_{t_i}I_T^2 - 2I_T I_i\circ\phi_{t_i} \\
&  ~~~~~~~~~~~~~~ + \big(I_i\circ\phi_{t_i}\big)^2|D\phi_{t_i}| \Big) \times M_i\circ\phi_{t_i} dy
\end{split}
\end{equation}
where $\times$ denotes scalar multiplication.

Taking the variation of the cost function with respect to $I_T$ gives
\begin{equation}
\begin{split}
\nabla_{I_T}E & = \sum_{i=0}^{N-1} \nabla_{I_T}h_i \\
& = \sum_{i=0}^{N-1} \frac{2}{\gamma^2}\big( |D\phi_{t_i}^{-1}|\circ\phi_{t_i}M_i\circ\phi_{t_i}I_T - I_i\circ\phi_{t_i}M_i\circ\phi_{t_i}\big) .
\end{split}
\end{equation}
Setting this gradient to zero and solving for $I_T$ gives a closed-form formula for updating template image in the coordinate system of $I_0$ at each iteration
\begin{equation}
I_T(y) = \frac{\sum_{i=0}^{N-1} (I_i M_i)\circ\phi_{t_i}(y)}{\sum_{i=0}^{N-1} (|D\phi_{t_i}^{-1}|M_i)\circ\phi_{t_i}(y) } .
\end{equation}

\subsection{Solving the Pullback Flow of Diffeomorphisms}
The pullback flow of diffeomorphisms $\phi^{-1}_t$ can be solved by the differential equation $\frac{d}{dt}\phi^{-1}_t = -D\phi^{-1}_t v_t$.
Suppose $\phi_t^{-1}$ and $v_t$ are discretized into $N$ computational-time points $1,2,\cdots,N$, denoted by $\phi^{-1}[1],\phi^{-1}[2],\cdots,\phi^{-1}[N]$ and $v[1],v[2],\cdots,v[N]$.
One approach to solve the above differential equation is to directly use the Euler method: $\frac{\phi^{-1}[i+1] - \phi^{-1}[i]}{\Delta_t} = -D\phi^{-1}[i]v[i]$.
We then obtain the following:
\begin{itemize}
	\item $\phi^{-1}[1] = Id$
	\item $\phi^{-1}[2] = \phi^{-1}[1] - \Delta_t(D\phi^{-1}[1])v[1] = Id - \Delta_tv[1]$
	\item $\phi^{-1}[3] = Id - \Delta_t(v[1]+v[2]) + \Delta_t^2 Dv[1]v[2]$
	\item $\phi^{-1}[4] = Id - \Delta_t(v[1]+v[2]+v[3]) + \Delta_t^2(Dv[1](v[2]+v[3]) + Dv[2]v[3]) -  \Delta_t^3 D(Dv[1]v[2])v[3]$
	\item $\cdots$
\end{itemize}
Since the computation of each $\phi^{-1}[i]$ involves high-order partial derivatives of the vector fields $v[1],v[2],\cdots,v[i-2]$, directly using the Euler method can cause high-frequency noises to the pullback flow.

We notice that solving the O.D.E $\frac{d}{dt}\phi_t = v_t(\phi_t)$ for the pushforward flow of diffeomorphisms using the Euler method only requires the first-order derivatives of the vector fields.
Therefore, we compute the Jacobian matrix of $\phi^{-1}_t$ using $D\phi^{-1}_t = \big((D\phi_t)\circ\phi^{-1}_t\big)^{-1}$ instead of first-order partial derivatives of $\phi^{-1}_t$.
Our modified approach for solving $\frac{d}{dt}\phi^{-1}_t = -D\phi^{-1}_t v_t$ becomes
\begin{equation}
\frac{\phi^{-1}[i+1] - \phi^{-1}[i]}{\Delta_t} = -\big((D\phi[i])\circ\phi^{-1}[i]\big)^{-1}v[i].
\end{equation}
This approach is more robust since it only requires first-order derivatives of the vector fields.

Directly using $J(\phi^{-1}[i])$ to compute the Jacobian determinant of $\phi^{-1}[i]$ involves computation of the second-order derivatives of the vector fields.
Therefore, we use a more robust approach
\begin{equation}
J(\phi^{-1}[i]) = \frac{1}{J(\phi[i])\circ\phi^{-1}[i]}.
\end{equation}

\subsection{Optimal Control and Pontryagin's Minimum Principle}
\label{sec:pmp}
The objective of optimal control theory is to find acceptable trajectories of the control signals that minimize (or maximize) the performance measure of a system satisfying certain constraints~\cite{donald1970}.
An optimal control problem usually consists of: (1) a set of differential equations that describes the flow of the state variable; (2) system constraints; (3) performance measure.
In this manuscript, we use the traditional control theory notation $x(t)$ for the state variable, $u(t)$ as the control variable, and $p(t)$ as the costate variable.
This may cause a little bit confusion with the rest of the manuscript in which we use $x$ as the spatial coordinate of image domain.
Suppose a system can be described by the following differential equation:
\[
\dot{x}(t) = a(x(t),u(t),t)
\]
where the function $a(\cdot,\cdot,\cdot)$ describes the physical process.

The goal of optimal control is to find an admissible control $u(t)$ that minimizes the following performance measure:~\cite{donald1970}
\begin{equation}
J(u) = \sum_{i=1}^{N}h_i(x(t_i),t_i) + \int_{t_0}^{t_N}g(x(t),u(t),t)dt
\end{equation}
where we assume that the initial time $t_0$, initial state $x_0$ and final time $t_f$ are fixed.

We introduce the costate variable $p(t)$ that has the same dimension as the state variable $x(t)$, and define the Hamiltonian:
\[
H(x(t),u(t),p(t),t) = g(x(t),u(t),t) + p(t)\cdot a(x(t),u(t),t)
\]
Pontryagin's Minimum Principle (PMP) gives necessary conditions for optimal solutions to the above control problem:~\cite{donald1970}
\begin{equation}
\begin{cases}
\dot{x}(t) = \frac{\partial H}{\partial p}(x(t),u(t),p(t),t) \\
\dot{p}(t) = -\frac{\partial H}{\partial x}(x(t),u(t),p(t),t)\\
0 = \frac{\partial H}{\partial u}(x(t),u(t),p(t),t)
\end{cases}
\label{eq:optimal_1}
\end{equation}
for all $t\in[t_0,t_N]$ except the time points $t_1,\cdots,t_{N-1}$, and also
\begin{equation}
\begin{cases}
p(t_N) = \frac{\partial h_N}{\partial x}(x(t_N),t_N) \\
p(t_i^{-}) = p(t_i^{+}) + \frac{\partial h_i}{\partial x}(x(t_i),t_i)
\end{cases}
\label{eq:optimal_2}
\end{equation}
for all $i\in\{1,2,\cdots,N-1\}$.

\subsection{L-BFGS Optimizer}

The limited-memory Broyden–Fletcher–Goldfarb–Shanno (L-BFGS)~\cite{liu1989limited} optimizer was used in the GDR algorithm to find the descent direction at each iteration.
L-BFGS is a quasi-Newton method that uses limited computer memory to estimate the descent direction, which is the product of the inverse of the Hessian matrix and the gradient vector.
Instead of explicitly computing and storing the Hessian matrix, L-BFGS computes the matrix-vector product by using only the past $m$ updates of the position vector $x_k$ and the gradient vector $g_k$.
In the GDR algorithm, $x_k = v_t^k$ and $g_k = v_t^k + K*\big(I^k(t)\nabla\lambda^k(t)\big)$, where the superscript $k$ denotes the $k$-th iteration.
The L-BFGS algorithm is stated in Algorithm~\ref{al:L-BFGS}.
\begin{algorithm}
	\SetKwInOut{Input}{Input}\SetKwInOut{Output}{Output}\SetKwInOut{Initialization}{Initialization}
	\Input{$x_k,\cdots,x_{k-m}$, $g_k,\cdots,g_{k-1}$}
	\Output{Z}
	\BlankLine
	$q = g_k$\;
	\For{$i = k - 1,\cdots, k -m$}
	{
		$s_i = x_{i+1} - x_i$\;
		$y_i = g_{i+1} - g_i$\;
		$\rho_i = \frac{1}{<y_i,s_i>}$\;
		$\alpha_i = \rho_i<s_i,q>$\;
		$q = q -\alpha_i y_i$\;
	}
	$\gamma_k = \frac{<s_{k-1},y_{k-1}>}{<y_{k-1},y_{k-1}>}$\;
	$Z = \gamma_k q$\;
	\For{$i = k -m, \cdots, k - 1$}
	{
	    $\beta_i = \rho_i <y_i, Z>$\;
	    $Z = Z + s_i(\alpha_i - \beta_i)$\;
	}
	Z = -Z
	\caption{The L-BFGS algorithm~\cite{liu1989limited}. This algorithm gives an estimate of the descent search direction based on the past $m$ location and gradient vectors. The operator $<,>$ computes the inner product between two velocity fields.}
	\label{al:L-BFGS}
\end{algorithm}  
We used the strong Wolfe condition~\cite{wolfe1969convergence} to find an acceptable step size $s$. 
Given a step size $s = 1$, $s$ is reduced by half until it satisfies the following inequalities
    \begin{align*}
    (1) & ~~ E(x_k + sZ) ~~ \leq ~~ E(x_k) + c_1 s <Z,g_k>\\
    (2) & ~~ |<Z,g_{k+1}>| ~~ \leq ~~ |c_2<Z,g_k>|
\end{align*}

where $Z$ is the descent direction, $E$ is cost function, and we chose  $c_1 = 10^{-4}, c_2 = 0.9$.

\subsection{Artifact Detection U-net Architecture}
\begin{figure}[!hbt]
	\centering
	\includegraphics[width=1.0\linewidth]{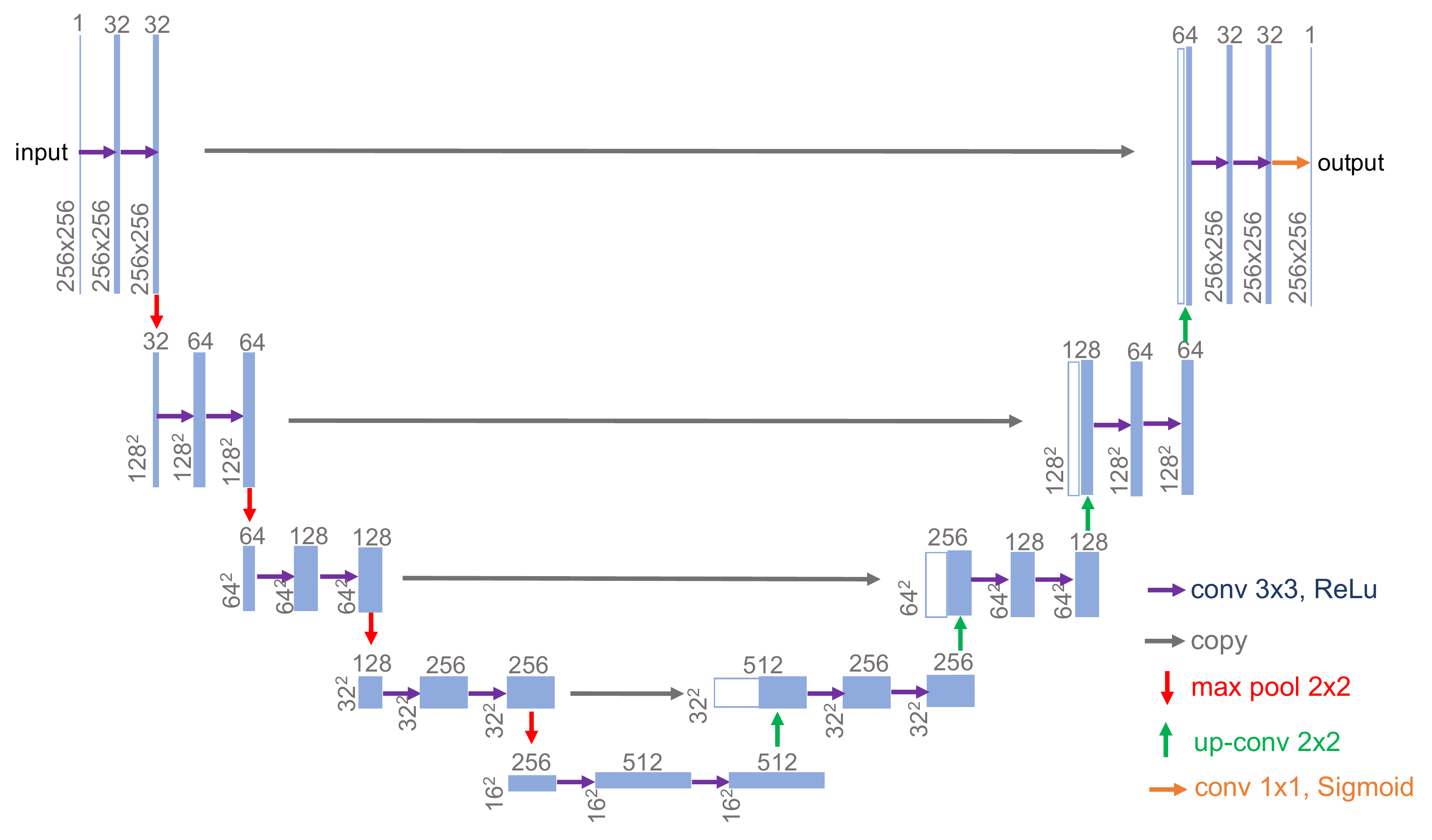}
	\caption{Architecture of the U-net model used for automatic detection of motion artifacts. 
	The height and width of the images are in the lower left edge of the boxes, and the number of feature maps is on the top of the box.
		\label{fig:unet}}
\end{figure}

Figure~\ref{fig:unet} shows the convolutional U-net~\cite{ronneberger2015u} architecture we used for automatic detection of artifact regions in 2D CT slices.
The encoder consists of repeated convolutional layers, each followed by a rectified linear unit (ReLU) and a max pooling layer.
The encoder reduces the the spatial resolution and increases the depth of the feature maps.
The decoder passes feature maps through a sequence of up-convolutions and concatenations with image features from the encoder followed by repeated convolutional layers.
The role of the decoder is to map the low-resolution feature maps from the encoder to high-resolution feature maps for pixel-wise image segmentation.
\label{sec:u-net}

\section*{Acknowledgments}
This work was supported by National Cancer Institute of the National Institutes of Health (NIH) under award numbers R01CA166703 and R01CA166119, National Heart, Lung, and Blood Institute (NHLBI) of NIH under award number R01HL142625,
the Department of Radiology at Stanford University, and Radiology Science Laboratory (Neuro) from the Department of Radiology at Stanford University.

\bibliographystyle{IEEEtran}
\bibliography{main.bib} 

\end{document}